\documentclass[twocolumn]{aastex631}

\usepackage{amsmath}
\usepackage{soul}

\newcommand\tess{\textit{TESS}}
\newcommand\MM{M\&M}
\newcommand\MMs{M\&Ms}

\submitjournal{ApJ}

\shorttitle{TESS \MM\ EBs}
\shortauthors{Oddo et al.}

\begin{document}


\title{A Catalog of \MM\ Eclipsing Binaries with TESS}

\correspondingauthor{Dominic Oddo}
\email{doddo@unm.edu}

\author[0000-0002-2702-7700]{Dominic Oddo}
\affiliation{Department of Physics and Astronomy, The University of New Mexico, 210 Yale Blvd NE, Albuquerque, NM 87106, USA}

\author[0000-0003-2313-467X]{Diana Dragomir}
\affiliation{Department of Physics and Astronomy, The University of New Mexico, 210 Yale Blvd NE, Albuquerque, NM 87106, USA}

\author[0000-0003-0501-2636]{Brian P. Powell}
\affiliation{NASA Goddard Space Flight Center, 8800 Greenbelt Road, Greenbelt, MD 20771, USA}

\author[0000-0001-9786-1031]{Veselin B. Kostov}
\affiliation{NASA Goddard Space Flight Center, Greenbelt, MD 20771, USA}
\affiliation{SETI Institute, 189 Bernardo Ave, Suite 200, Mountain View, CA 94043, USA}

\author[0000-0002-7127-7643]{Te Han}
\affiliation{Department of Physics \& Astronomy, The University of California, Irvine, Irvine, CA 92697, USA}

\begin{abstract}

We present a catalog of 1292 low-mass (\MM) short-period eclipsing binaries observed by the \tess\ mission. Eclipsing binaries are useful for many aspects of stellar astrophysics, including calibrating stellar models, and catalogs of eclipsing binary properties provide context for population-level inferences. In this work, we present our selection criteria for our catalog, along with steps taken to characterize the orbital and physical properties of our EBs. We further detail crucial steps in vetting our catalog to ensure that the systems in our catalog have a high likelihood of being true \MMs. Our sample consists primarily of binaries with short-period, circular orbits and often manifest as high-mass ratio ``twin'' \MMs. We find that our sample of \MMs\ does not exhibit the same excess of contact binaries that similar samples of solar-type binaries do. Using a novel statistical approach, we find a tidal circularization period of approximately 2.0 days for these binaries, which is significantly shorter than the value typically given in the literature of 7 days. Finally, we explore the prospects for additional companions to our \MMs. Future avenues include spectroscopic follow-up of bright \MMs\ to better-characterize low-mass stellar properties and a deeper assessment of the presence of tertiary companions.

\end{abstract}

\section{Introduction} \label{sec:intro}

Eclipsing binaries (EBs) are important to stellar astrophysics for a variety of reasons. Well-characterized EBs can provide constraints to fundamental stellar properties (e.g. \citet{Kraus_2011,Parsons_2018,Swayne_2024}). They can also serve as excellent probes of stellar activity (e.g. \citet{Huang_2020,Sethi_2024}), as well as stellar formation (e.g. \citet{Bate_2015,Offner_2023}). This in turn can be used to fine-tune models of stellar properties, which are in high demand in astrophysics. In particular, a detailed knowledge of extrasolar planet host stars directly correlates with more detailed information about the planets themselves, which is leading to increasingly precise knowledge about the formation and evolution processes of planets en masse (e.g. \citet{VanEylen_2018,Berger_2020,Sullivan_2024}). 

Stellar multiplicity rate (i.e. the proportion of systems in binaries or higher-order systems) negatively correlates with primary mass, where lower-mass stars tend to be in binaries less frequently than their higher-mass counterparts. This may be indicative of the formation environments of binary stars, where high-density regions are likely to collapse on shorter timescales into larger protostellar cores, whose gravitational influences lead to a likelier collapse of high-mass neighbors \citep{Beuther_2025}. Well-studied M dwarf binaries, like the CM Draconis system (e.g. \citet{Jenkins_CMDrac_1996,Martin_CMDrac_2024}) are therefore important due to their relative rarity. Since M dwarfs are dimmer at a given distance than stars of earlier spectral type, we are limited to M dwarfs in our local neighborhood in the Galaxy, compounding uncertainties relating to galactic stellar properties. 

The NASA \tess\ mission \citep{Ricker_2015}, launched in 2018, is optimized to search for transiting exoplanets orbiting bright, nearby stars, thus offering a census of short-period planets in our backyard. Because of its high sampling cadence and near-complete sky coverage, it is also an excellent mission to study short-period eclipsing binary systems (P $\lesssim$ 10 d). \tess's observing strategy is such that most sources in its field are only observed for $\sim27$ days at a time (with the notable exception of the continuous viewing zones - CVZs -  around the ecliptic poles). This means \tess\ is most sensitive to binaries with the shortest orbital periods. This is due to the high probability of such binaries to eclipse. In reality, the peak of the overall binary separation distribution is much further beyond \tess's sensitivity limits. \tess\ has already demonstrated the ability to yield a large number of short-period EBs \citep{Prsa_2022,IJspeert_2024,Kostov_2025} thanks to its large sky coverage. 

There are three major mechanisms to describe the formation of tight binaries: 1. turbulent fragmentation; 2. disk instabilities; 3. capture \citep{Offner_2023}. Both turbulent fragmentation (either in cores or filamentary structures) and disk instabilities tend not to form companions with low mass ratios. Turbulent fragmentation tends to yield binaries with separations on the scale of hundreds of AU, whereas disk instabilities tend to yield binaries that are separated by tens to a couple hundred AU. Capture might yield comparatively wider distributions of mass ratios and separations, but the probability of capture events is low, and thus capture is likely responsible for only a small portion of binaries    \citep{Bate_2015}. Population-level insights into the primary formation channels of short-period binary systems may be revealed by statistical evaluation of their bulk properties, such as eccentricities and mass ratios. There have been many studies exhibiting a statistically-significant excess of so-called "twin" binaries (those with mass ratios $q > 0.95$; \citet{Raghavan_2010,ElBadry_2019,Winters_2019,Matson_2025}). With this catalog, we seek to explore these properties for a large number of short-period, low-mass EBs, thus adding to the canon of eclipsing binaries in the overall portrait of stars, stellar properties, and stellar astrophysics.

Short-period binaries tell an extreme story about binary evolution because large amounts of angular momentum must be removed, and they are relatively rare. Indeed, the close binary fraction for M dwarfs is generally found to be relatively low at $1.8\pm1.8$\% \citep{Fischer_1992} from 0.04-0.4 AU separations, $3-4$\% within 0.4 AU \citep{Clark_2012}, and $11_{-4}^{+2}$\% within 90 days \citep{Shan_2015} (although \citet{Shan_2015} express that their value might be a slight over-estimate if some of their candidates are not verified). Short-period binaries might be formed via multi-core fragmentation \citep{Bonnell_1994,Bate_2000} or via instabilities in a rotating disk (i.e. disk instabilities, \citet{Kratter_2010}). 
Migration and accretion also play crucial roles. Simulations incorporating both turbulent and radiative feedback have shown that low-mass binaries likely form via turbulent fragmentation \citep{Offner_2010} in bound filamentary structures, which facilitate migration over $10^5$ yr timescales \citep{Offner_2016,Lee_2019}. \citet{Moe_2018} found that short-period solar-type binaries must have migrated during the early pre-main sequence (MS) phase of formation. Differential accretion may equalize the masses of components in short-period binaries \citep{Bate_2002}, especially so for low-mass short-period binaries \citep{Nefs_2013}. Further orbital evolution may occur due to interactions with a circumbinary disk, including driving binary eccentricities to certain values depending on binary and disk properties \citep{Artymowicz_1991,Siwek_2023,Valli_2024}.

It is important to examine short-period binaries in the context of hierarchical systems and tertiary companions. Another evolution channel which could yield short-period binaries involves hardening by dynamical Kozai-Lidov cycles \citep{Eggleton_2006,Fabrycky_2007}, although the observed orbital properties of a majority ($\sim60$\%) of solar-type binaries are not consistent with this mechanism \citep{Moe_2018}. Nevertheless, many short-period binaries are expected to have tertiary companions \citep{Delgado-Donate_2004}. Searches for tertiary companions to binaries have yielded many results with common proper motion analysis \citep{Allen_2012}, high-resolution imaging \citep{Tokovinin_2006,Powell_2023}, and eclipse-timing analysis \citep{Borkovits_2016,Hanjdu_2022_OGLE,Hajdu_2022_CoRoT,Mitnyan_2024,Borkovits_2025}. Broadly speaking, these results demonstrate that a very large proportion of short-period binaries are expected to have tertiary companions.

The remainder of the paper is structured as follows: In Section \ref{sec:sample_data}, we detail our sample selection and data used to compile this catalog. Broadly speaking, we discuss our methodology in Section \ref{sec:meth}, including our ephemeris search and validation in Sec. \ref{subsec:ephem}, our vetting in Sec. \ref{subsec:vet}, and our spectral energy density fitting in Sec. \ref{subsec:sed_fit}. We detail our results in Section \ref{sec:results}, followed by a discussion of orbital and physical properties in Section \ref{sec:disc}. We offer prospects for future work and conclude in Section \ref{sec:conc}.

\section{Sample Selection and Data} \label{sec:sample_data}

\begin{deluxetable*}{cccccccccccc}

\rotate
\tablecaption{Observables \label{tab:observables}}

\tablehead{\colhead{TIC} & \colhead{Gaia DR 3 ID} & \colhead{RA} & \colhead{Dec} & \colhead{TESS mag} & \colhead{GAIAmag} & \colhead{H mag} & \colhead{Contratio} & \colhead{High contam} & \colhead{Distance} & \colhead{RUWE} & \colhead{TESS Sectors} \\ 
\colhead{} & \colhead{} & \colhead{(deg)} & \colhead{(deg)} & \colhead{} & \colhead{} & \colhead{} & \colhead{} & \colhead{(boolean)} & \colhead{(pc)} & \colhead{} & \colhead{}} 

\startdata
316050212 & 1439709463238102400 & 7.478886 & 63.554681 & 14.8546 & 15.8736 & 12.622 &  & False & 242.9237 & 0.9762 & 18 \\
381805454 & 2260823166312415232 & 238.017114 & 42.32966 & 14.9306 & 16.0272 & 12.902 & 0.223103732 & False & 501.7395 & 1.0297 & 18 \\
61224047 & 2036874569262319616 & 242.593732 & 33.964518 & 14.7284 & 15.8032 & 12.757 & 0.000034873512 & False & 308.5495 & 0.9737 & 11 \\
186510259 & 2235320131549037824 & 238.07689 & 30.592245 & 14.4302 & 15.3576 & 12.388 &  & False & 222.355 & 1.0758 & 17 \\
176790779 & 1636502837516468864 & 358.515371 & 57.289926 & 14.8704 & 15.9278 & 12.848 & 0.7003239 & False & 213.9722 & 1.0732 & 17 \\
464343363 & 2144147356061898752 & 256.38628 & 44.912608 & 14.5751 & 15.6027 & 12.56 & 0.006342842 & False & 319.7981 & 1.4969 & 13 \\
25681445 & 2126014141581543424 & 340.15148 & 59.258317 & 14.9914 & 16.1444 & 12.933 & 1.00003588 & True & 315.7598 & 1.0529 & 17 \\
196902457 & 2021944129112694656 & 260.686271 & 45.107209 & 14.7058 & 15.9574 & 12.726 & 0.5215174 & False & 262.655 & 1.0402 & 13 \\
55108492 & 2149876803777511808 & 2.304055 & 72.317972 & 13.9345 & 15.1191 & 11.668 & 0.0337143764 & False & 313.1169 & 1.3733 & 9 \\
267806373 & 2158790651919783552 & 234.553934 & 25.957777 & 14.5436 & 15.8139 & 12.634 & 0.0236891154 & False & 168.6184 & 0.9584 & 17 \\
105767200 & 2104539545617061632 & 249.10292 & 43.37865 & 14.8092 & 16.0027 & 12.127 &  & False & 343.4384 & 1.0363 & 9 \\
273263945 & 2260552652093614080 & 254.602884 & 46.658273 & 14.8522 & 15.8186 & 12.635 &  & False & 216.8651 & 1.0332 & 19 \\
459083368 & 2079336952408224512 & 232.111277 & 20.435243 & 14.4848 & 15.8778 & 12.06 & 0.0005575706 & False & 223.5188 & 1.2808 & 19 \\
238879024 & 1422633600982835200 & 242.704432 & 41.574363 & 13.4772 & 14.5409 & 11.437 & 0.3765679 & False & 181.1113 & 1.017 & 19 \\
27302173 & 2019686866100324224 & 246.774423 & 51.651313 & 14.2617 & 15.6718 & 11.899 & 0.014906575 & False & 159.2136 & 2.008 & 13 \\
66849985 & 1833273671831508608 & 230.635148 & 37.072514 & 14.5739 & 15.6083 & 12.596 & 0.0102785826 & False & 226.9114 & 1.0319 & 19 \\
339253436 & 2161024893906539136 & 251.265504 & 47.2369 & 14.2526 & 15.3864 & 12.012 & 0.00558095472 & False & 117.4177 & 1.0882 & 19 \\
244208214 & 2054967548493258624 & 252.58636 & 46.65038 & 13.3322 & 14.5665 & 11.137 & 0.0536392964 & False & 352.4694 & 1.0131 & 5 \\
280623400 & 2289872435318491904 & 296.871612 & 21.897526 & 13.2042 & 14.2781 & 11.422 & 0.441232532 & False & 209.045 & 1.2663 & 18, 19, 25 \\
276105676 & 2160989056699546112 & 303.734833 & 24.414596 & 13.1133 & 14.0267 & 11.118 &  & False & 253.1537 & 0.9898 & 16, 17, 24 \\
\enddata

\tablecomments{Table 1 is published in its entirety in the machine-readable format.
      A portion is shown here for guidance regarding its form and content.}

\end{deluxetable*}

\subsection{Sample selection}\label{subsec:sample}
The starting point for this catalog was a list of 498,046 EB candidates, which were originally described in \citet{Powell_2021}. Briefly, the team generated millions of \tess\ Full Frame Image (FFI) light curves from the Primary Mission (first 2 years of observations) to a depth of T $< 15$ mags through a parallelized implementation of \texttt{eleanor} \citep{Feinstein_2019}, which is a Python module used primarily for the production of \tess\ light curves from FFIs, on the \textit{Discover} supercomputer at the NASA Center for Climate Simulation (NCCS). \citet{Kostov_2025} searched for eclipse-like features in \tess\ light curves using a neural network classifier, described further in \citet{Powell_2021}. Additional details about the neural network are described in \citet{Kostov_2025}, but it is important to note that a very high proportion of sources initially identified are not EBs ($56-86\%$), and it is very difficult to assess the completeness. This demonstrates the importance of taking steps to vet, verify, and calculate the properties of EBs, as \citet{Kostov_2025} did for the sub-sample of 10,001 EBs and as we do in this work. \citet{Kostov_2025} perform sensitivity tests of their sample by comparing the findings from the NN to the known Kepler and \tess\ EB catalogs (their Figs. 4-6). While they note deficiencies in their classifier compared to the Kepler EB catalog, their catalog appears to be mostly complete for \tess\ EBs across orbital periods to which \tess\ is sensitive. Further work to survey the completeness of this catalog, or the completeness of \tess\ in its identification of EBs is needed but beyond the scope of a catalog paper. 

We queried for parameters from the TESS Input Catalog \citep{Stassun_2019} for all of these TICs. Since we were interested in identifying low-mass \MM\ eclipsing binaries, we made an initial cut by selecting those TICs which were listed as having effective temperatures (T$_{eff}$) below 4,200 Kelvin (\texttt{T$_{eff}$ < 4200 K}). Although it is known that the TIC parameters are biased for EBs \citep{Stassun_2019}, this temperature cut was chosen to account for the fact that a secondary will typically contribute primarily in wavelengths redder than the primary star, and will not substantially increase the calculated T$_{eff}$. The temperature cut significantly reduced the size of our sample to 22,351 EB candidates.

\begin{figure}
    \centering
    \includegraphics[width=0.9\linewidth]{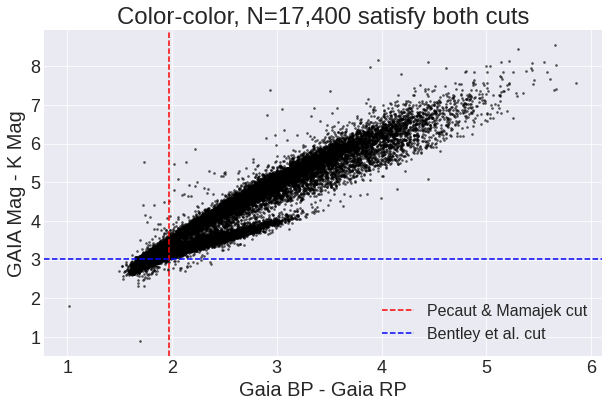}
    \caption{Gaia BP-RP vs Gaia G-K color-color plot for our candidate \MMs. Candidates falling above the blue line and to the right of the red line were then further analyzed.}
    \label{fig:color_cut}
\end{figure}

We performed color-color cuts to ensure that our candidates would be low-mass binaries, as shown in Fig. \ref{fig:color_cut}. We used color cuts in both Gaia BP-RP color from \citet{Pecaut_2013} \footnote{and subsequently updated by E. Mamajek in 2022 at \url{https://www.pas.rochester.edu/~emamajek/EEM_dwarf_UBVIJHK_colors_Teff.txt}} at $B_P-R_P > 1.97$  and in Gaia G - 2MASS K from \citet{Bentley_2019} at $G-K > 3.013$, which are the thresholds for an M0 star in the respective references. These selections were made based on the robustness of the measurements and the broad availability of these magnitudes for a large proportion of our EB candidates. 

We can examine the prospects for hidden post-main sequence companions such as white dwarfs to our M-dwarfs by checking the effectiveness of these color cuts. Typical single white dwarfs have Gaia BP-RP colors that are bluer than our color cuts, eliminating most white dwarfs \citep{Gentile_Fusillo_2021}. A white dwarf-main sequence (WD-MS) binary likely would form via mass transfer, wherein the presence of an M-dwarf companion to a white dwarf progenitor would cause the white dwarf progenitor’s outer layers to be stripped, possibly leading to a common envelope phase. In the case that a WD-MS binary is formed, the white dwarf will initially be very hot (tens of thousands of Kelvin), which would dominate the photometric output and thus affect the apparent color of the system. While the white dwarf is still hot, it will pull the color of the system bluer than our color cuts. Based on current white dwarf cooling models, it would take several billion years for the white dwarf to cool below a flux threshold that would not be photometrically detectable \citep{Lam_2022,Nayak_2024}. Because of this, we find it unlikely that a significant portion of our sources are WD-MS pairs, but it is a slim possibility.

We can also calculate the lower limit for the temperature of a white dwarf that would not be detectable by using the flux relation:

\begin{equation}
    \frac{F_{WD}}{F_{dM}} = \frac{R_{WD}}{R_{dM}}^2 * \frac{T_{WD}}{T_{dM}}^4
\end{equation}

Where subscripts WD and dM are White Dwarf and M-dwarf, respectively, and symbols F,R, and T are flux, radius, and temperature, respectively. By rearranging this equation and solving for the white dwarf temperature, we can calculate this limit. We must make an assumption of the “tolerable” amount of flux contribution from the white dwarf in this case. We can start by choosing a 1\% flux contribution ($\frac{F_{WD}}{F_{dM}} = 0.01$), which is within the bounds of photometric uncertainty and would likely not be detectable. For a “typical” M-dwarf in our catalog, with properties of $R_{dM} = 0.45$ R$_{sun}$ and $T_{dM} = 3500$ K, a typical white dwarf with a size of $R_{WD} = 0.02$ R$_{sun}$ with a temperature of 5250 K would equate to a 1\% flux contribution. This is not negligible, as there are white dwarfs that are cooler than this. However, from the above discussion and by performing this calculation, we have shown that the likelihood of a white dwarf companion in our sample is low.

Of the 22,351 candidates from above, 17,400 are redder than both of the color cuts, as shown in Fig. \ref{fig:color_cut}. Of these, we had light curves for 17,255 EB candidates, as described in \ref{subsec:tglc}. Binaries of earlier spectral type than M could still sneak through at this stage if they are reddened by interstellar dust, but this is accounted for during our spectral energy distribution fitting in Section \ref{subsec:sed_fit}. These were the initial EB candidates which were then analyzed further, including the ephemeris identification, on-target vetting, and light curve and spectral energy density (SED) fitting. 

Table \ref{tab:observables} shows observable properties for sources in our catalog from both the TIC and Gaia Catalog. From the TIC, we obtained Right Ascension and Declination (RA and Dec, respectively), magnitudes, and \tess\ contamination ratios (listed in our table as \texttt{contratio}). We delineate sources with high contamination with the flag \texttt{High Contam} in Table \ref{tab:observables} as those with contamination ratios greater than 1. From the Gaia Catalog, we obtained parallactic distances and Renormalized Unit Weight Error (RUWE), which is described in greater detail below. These are shown in Table \ref{tab:observables}.

\subsection{Observations and Data} \label{subsec:data}

\subsubsection{\tess} \label{subsec:tess}

\begin{figure}
    \centering
    \includegraphics[width=0.95\linewidth]{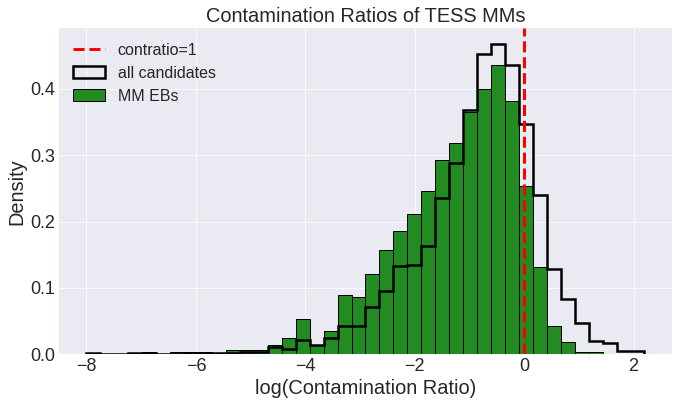}
    \caption{Flux contamination ratios, which is the ratio of the total contaminant flux to the flux of the target source, for all 22,351 TICs in our initial sample (outlined) and for our 1,292 \MMs, scaled to the same vertical axis for comparison. Note that the "Contamination Ratio" axis is in log scale. The \MMs\ are shifted left in this figure compared to the overall sample, indicating they are generally less contaminated.}
    \label{fig:cont_ratios}
\end{figure}

The Transiting Exoplanet Survey Satellite (\tess) is a space-based observatory designed to discover exoplanets transiting bright, nearby stars through a nearly all-sky photometric survey \citep{Ricker_2015}. In this work, we utilize light curves generated from \tess\ Full Frame Images (FFIs) from the Primary Mission (PM; Sectors 1-26) and First Extended Mission (EM1; Sectors 27-55).

\tess\ has large pixels, which will cause flux from nearby stars to contaminate the light curves of sources in our sample. In Fig. \ref{fig:cont_ratios}, we plot the contamination ratio of all 22,351 sources in our initial \MM\ candidate sample as well as the contamination ratios of the 1,292 sources in the final catalog queried from the TIC. See \citet{Stassun_2019} Section 3.2.1 for details regarding the method for calculating flux ratio, but contamination ratio is defined as the ratio of contaminant flux to the flux of the source, setting the bounds from 0 to any positive number. This calculation was performed having made certain assumptions about how the aperture photometry would be extracted, but we would expect that the actual contamination ratios for our TGLC light curves (see below) are lower. While Fig. \ref{fig:cont_ratios} shows that many of our sources have contamination ratios that are quite low, there is a sub-population for which contamination can be quite high, delineated by the vertical line at a contamination ratio of 1. In Table \ref{tab:observables}, we mark sources with high contamination ratios (greater than 1.0). This is an arbitrary threshold, and different thresholds should be chosen for different purposes. High contamination will principally affect the eclipse depths.

\subsubsection{TGLC light curve extraction}\label{subsec:tglc}

\begin{figure}
    \centering
    \includegraphics[width=0.9\linewidth]{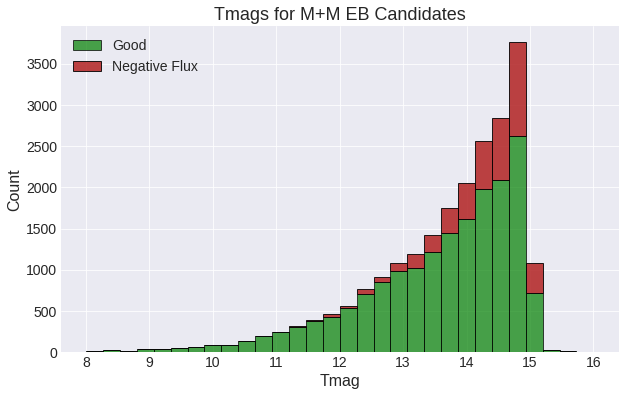}
    \caption{Stacked histogram of \tess\ band magnitudes of the initial 22,351 M\&M candidates in our sample, including light curves for which there were negative and no negative (``good") flux values. The distribution shows a brightness-limited sample with a cutoff at T $\approx 15$ mag.}
    \label{fig:Tmag_dist}
\end{figure}

There are many ways to extract light curves from \tess\ images. The goal is to minimize contributions from instrumental and background flux while preserving signals which are astrophysical in nature to the best degree possible. 



TESS-Gaia Light Curve \citep[TGLC]{Han_2023} is a relatively new method for extracting \tess\ light curves. The code utilizes an effective Point Spread Function (PSF) to remove background contamination and is effective even for dim sources in crowded fields. TGLC does this by leveraging prior knowledge from the Gaia mission \citep{Gaia_2016,Gaia_2021} to map known sources. 
It then applies aperture photometry to the contamination-removed images and produces the TGLC aperture light curve, which we use in our study. We do not use the TGLC PSF light curves because they apply a second model of PSF focusing on removing variable contaminations, which do not apply to most of our systems. One of the advantages of using TGLC is that light curves are available to a depth of 15th \tess\ magnitude, which greatly increases our initial sample size relative to other light curve extraction methods.

\begin{figure}
    \centering
    \includegraphics[width=0.95\linewidth]{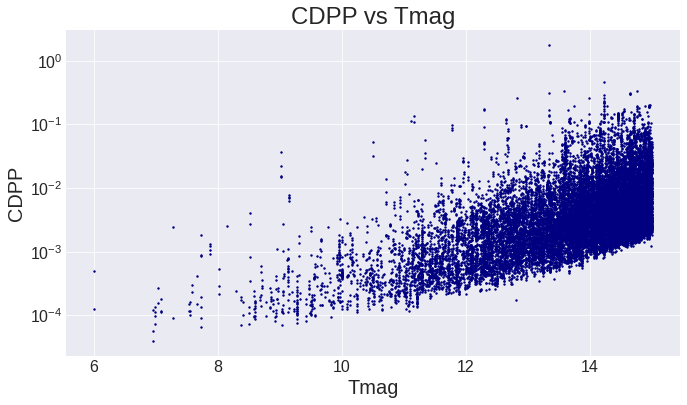}
    \caption{\tess\ band magnitudes vs Combined Differential Photometric Precision (CDPP) for our \texttt{tglc} light curves. The general expectation that noise will increase for dimmer systems is met.}
    \label{fig:Tmag_CDPP}
\end{figure}

Out of the 17,400 sources which pass, we successfully cross-matched 17,255 sources. We downloaded all available light curves up to \tess\ sector 55 for all 17,255 sources. We masked out points during spacecraft momentum dumps, which constitute a handful of points roughly every 3 days during the Primary Mission in order to maintain a reasonable noise level in the momentum wheels. Through visual inspection, we saw no evidence that this biased our determination of the orbital periods of our EBs. We also masked out points that were flagged with quality flags \citep{Twicken_2020}. 

We identified 4,271 sources for which there were negative fluxes in the light curves out of our 17,255 sources with \texttt{tglc} light curves. Negative fluxes occur when the background flux that is removed is greater than the brightness of the target itself, which most often occurred during eclipse due to the variability. This would indicate that the signal is off-target, and the removal of negative-flux light curves serves as a natural way to remove off-target signals. In the case that the negative fluxes occurred during binary eclipses, this would indicate that the true binary signal in the light curve was actually caused by a different source off-target. Fig. \ref{fig:Tmag_dist} shows the stacked histogram of "good" sources and sources for which negative fluxes were detected in the light curve (the top of each bin represents the total of both "Good" and "Negative Flux" in the bin). We removed all sources with any negative fluxes at this stage.

To get a sense for the noise properties of our light curves, we calculated the Combined Differential Photometric Precision (CDPP), as described by \citet{Christiansen_2012}. To do this, we masked the binary eclipses according to the ephemerides identified in Sec. \ref{subsec:ephem} and then performed flattening with the \texttt{wotan} Tukey-biweight algorithm \citep{Hippke_2019}. 
Fig. \ref{fig:Tmag_CDPP} shows the calculated CDPP for our light curves as a function of \tess\ magnitude after these steps. Despite the eclipse-masking and flattening steps, there is still a significant amount of vertical scatter in the magnitude-CDPP plot, which may be due to the inability for these simple steps to fully remove all variability signals. Generally speaking, the expectation that CDPP increases for dimmer sources (i.e. dimmer sources exhibit more noise) is met.

\section{Methodology}\label{sec:meth}
\subsection{Ephemeris Search and Validation} \label{subsec:ephem}

\subsubsection{Period Identification and Eclipse Characterization}\label{subsec:per_id}
Utilizing the list of 12,984 candidate \MMs, we performed an ephemeris search to identify the binary periods, times of eclipses, and eclipse depths and durations. All code described in this section was applied uniformly to all sources whose light curves did not contain negative flux values. 

We performed an initial period search by running a box least-squares (BLS) periodogram with varying box durations and a heuristically-determined set of periods, which is calculated based on the data cadence and duration. Box durations ranged from a minimum of 0.04 days to a maximum of 0.34 days. For each periodogram, we identified the 5 peaks with highest power and their associated durations. We then iterated through candidate periods, ranked by periodogram power, and performed the following steps:
\begin{enumerate}
    \item We calculated the best-fit box model for each of these peaks, including the time of mid-eclipse and eclipse depths and durations.
    \item  We then phase-folded the light curve to the candidate period and binned the phase-folded points - to account for noise - and found the minimum outside of the eclipse duration. This represented the location of the secondary eclipse, which we also calculated the depth and duration of.  We required the primary eclipse to be deeper than the secondary eclipse by definition; if this was not the case, our code flipped the original phases of the primary and secondary eclipses.
    \item Using the candidate ephemeris, we counted the number of primary and secondary eclipses in the light curve (not including gaps) to ensure periodicity of the signal; we required at least two primary eclipses and one secondary eclipse or two secondary and one primary eclipse to confirm a period.
    \item We calculated the depths of the primary and secondary eclipses on a sector-by-sector basis. The reason for this was to account for possible variation in the eclipse depths due to varying noise levels in different sectors; i.e. sometimes the depth of the eclipse was different in the PM sector(s) than it was in the EM sector(s). We calculated the coefficient of variation (CV; ratio of standard deviation to mean of the eclipse depths in the sector) of the eclipse depths when there was more than one eclipse in a given sector using $CV = \frac{\sigma_{\textup{ eclipse depths}}}{\langle\textup{eclipse depths}\rangle}$. This was done to ensure consistency of the signal within any given sector and was a robust way of accounting for the periodogram attempting to fit noise in the light curve. If the eclipse depths within a given sector were not consistent, the candidate period was not passed. 
    \item We then masked the eclipses according to the ephemeris of the candidate signal and calculated the out-of-eclipse CDPP for the candidate period. This step was to check goodness of fit for the candidate ephemeris.
    \item For each candidate ephemeris, we assigned a score $S$ which was calculated from normalized periodogram power, normalized CV, and normalized CDPP according to the following equation:

    \begin{equation}
        S = \frac{y(P)-min(y)}{max(y)-min(y)}*\frac{1}{CDPP}*\frac{1}{CV}
    \end{equation}

    where $y$ is periodogram power, so $y(P)$ is the power of the periodogram at the candidate period, $min(y)$, $max(y)$ are the minimum and maximum values of the periodogram power for a given light curve, respectively, and CV as in Step 4.

    \item We chose the best score out of the 5 candidate periods for a given source. Generally, the first or second peaks in the periodogram were the final periods from the catalog. 

\end{enumerate}  

We applied this period search framework to each of the candidate EBs and found a period in 97.4\% of cases. In the few cases where a period was not successfully identified, there were no visible eclipses identified. 

\subsubsection{Ephemeris Validation}\label{subsec:ephem_val}

We visually inspected light curves for all of our sources that passed the previous stage as a check of the effectiveness of the period identification. This accounted for multiple types of unintended behavior, including 1. instances where a harmonic of the true period was identified by the code (e.g. 2X or $\frac{1}{2}$X the true period), 2. false positives (e.g. the BLS triggering on noise), 3. incorrect secondary phase location. To this end, we built a custom light curve viewer which phase-folded to the period identified, highlighted in-eclipse points, and showed the eclipses in isolation. In a small number of cases, this led us to eliminate the candidate EB from the final catalog. Some reasons for this included 1. excessive noise in the light curve, thus obscuring any periodicity, 2. not enough eclipses to confirm periodicity (e.g. only one primary and one secondary, or just a single eclipse signal), 3. the failure to identify any reliable period at all, e.g. the periodicity algorithm triggering on noise with no other signals visible in the light curve. In a small number of cases, this visual inspection prompted us to re-run the BLS periodogram in an attempt to recover the true period. From there, we inspected the candidate periods individually and chose the correct period based on visual inspection of the light curve, with preferential treatment for periods which had high power in the periodogram. 

Visual inspection is a reliable method to check the results of the ephemeris search code for a tractable number of EBs. We were able to verify the success of the method in calculating properties such as secondary phase location and depth, Further, we were able to account for and correct errors, such as if the eclipse duration was improperly calculated. 

Such a visual examination will introduce a bias in our sample against signals which are not verifiable by eye. While the stacking of shallow eclipses in phase space meant that many such signals could be identified, we are still biased against binaries with very low mass ratios (and those that are inclined relative to the line of sight with marginally grazing eclipses). 

\subsection{Vetting}\label{subsec:vet}

We performed flux-level vetting to ensure that candidate signals are astrophysical and not instrumental in nature. To this end, we used the \texttt{LEO-vetter} \citep{LEO_2024} code\footnote{\url{https://github.com/mkunimoto/LEO-vetter/tree/main}}. While \texttt{LEO} is nominally designed for vetting periodic exoplanet signals, we modified the tests that are usually used to qualify a ``passed'' signal in the base case. We disabled the typical tests that filter out EBs, including the V-shaped eclipse check, the depth check (to determine whether the candidate signal would be planetary in nature), and the sinusoidal variation check.

\tess\ cameras have quite large pixel sizes, with pixels approximately 21 arcseconds across. This leads to a relatively high chance of signals in a light curve being due to blends from nearby stars, and necessitates checking whether a given periodic signal is on target. For the sources that passed flux-level vetting, we used \texttt{LEO} to create difference images of in-eclipse and out-of-eclipse times, thus localizing the eclipse on a pixel image. This technique is built atop a framework that was originally developed for the \emph{Kepler} mission and has been adapted for the \tess\ mission (see \citet{Bryson_2013} for more details). By cross-matching against a more precise localization for the target source from the Gaia catalog, we determined the angular distance (or ``offset'') between the source and the candidate binary signal. 
We set an angular offset threshold at 20 arcseconds, where a source was classified as ``on-target'' if the offset was less than this. By choosing a threshold which is relatively high, we can robustly account for statistical uncertainties in the offset calculation. This may have led to a small number of off-target sources in our catalog, but this method is still reliable at large scale. 

\begin{figure}
    \centering
    \includegraphics[width=0.95\linewidth]{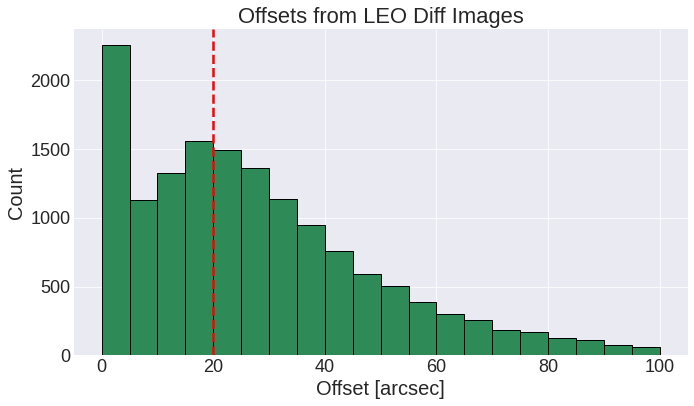}
    \caption{Histogram of offsets between the source position and the localization of the eclipse signal. We set the threshold for a source as "on-target" if the offset was less than 20 arcsec.}
    \label{fig:LEO_offset}
\end{figure}

As shown in Fig. \ref{fig:LEO_offset}, a large proportion of our sources were calculated to have high systematic offsets between the source and eclipse localization positions. Thus, this step eliminated a large fraction of our sources. This indicates the importance of performing vetting, as such a large proportion of off-target sources would surely bias our sample away from low-mass binaries. This further indicates that while a large portion of the excluded sources are likely to be real EBs, they cannot be classified as low-mass \MM\ EBs and are thus out of the scope of this paper. Later work could include reclassifying or re-examining these excluded sources in future EB catalogs.

We also used this step to de-duplicate and eliminate blends. We cross-checked our catalog to search for sources with similar ephemerides (i.e. similar periods and mid-eclipse times) and proximal sky locations, which would indicate that these sources are blended. We used the summary plots generated by \texttt{LEO} to identify the correct source for a given EB signal. In a small handful of cases, the EB signal could not be confidently matched to any source in our catalog. These were instances where the signal was more likely matched to a binary pair of earlier spectral type than M-dwarf, in which case the signal was eliminated from our catalog. 

\subsection{SED fitting}\label{subsec:sed_fit}

\begin{figure}
    \centering
    \includegraphics[width=0.95\linewidth]{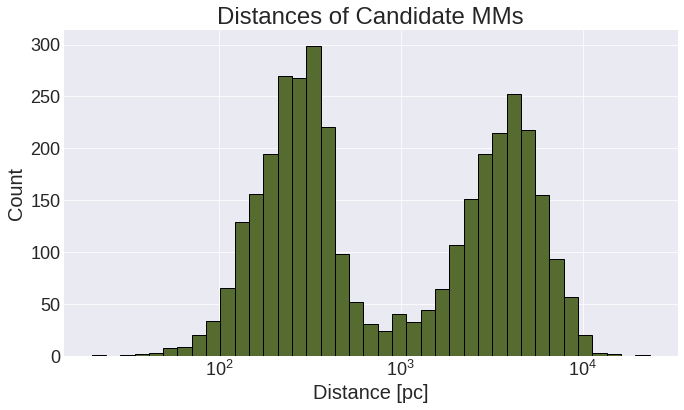}
    \caption{Gaia DR3 distances of candidate MMs which passed to the SED fitting phase. Note that distance (in parsecs) is in log space.}
    \label{fig:all_dists}
\end{figure}

Using the orbital properties we calculated in Sec. \ref{subsec:ephem_val} for the candidates that were shown to be on target by our vetting in Sec. \ref{subsec:vet}, we then moved to calculate physical component properties for the stars in our binaries. We used the publicly-available \texttt{SEDFit} code\footnote{\url{https://github.com/mkounkel/SEDFit/tree/main}} \citep{sedfit}. \texttt{SEDFit} is a Spectral Energy Density (SED) fitting routine which fits broadband photometric magnitudes that have been passband-corrected for dust reddening to stellar atmospheric models using parallactic distances from Gaia DR3. The magnitudes used include low resolution XP spectrum from Gaia, which spans the range in wavelength of 336–1020 nm, as well as the fluxes in Johnson and Cousins filters, along with fluxes from the Two Micron All-Sky Survey (2MASS), Wide-field Infrared Survey Explorer (WISE), Sloan Digital Sky Survey (SDSS), Gaia, and Galaxy Evolution Explorer (GALEX) surveys. We used the default \textbf{BT-Settl} model atmospheric grids \citep{Allard_2011}. In Fig. \ref{fig:all_dists}, we show Gaia DR3 distances for the candidate \MM\ EBs which passed to this SED fitting phase. We note that because of our color cut for redness (shown in Fig. \ref{fig:color_cut}), the more distant EB candidates were found by the SED fit to be binaries of earlier spectral type which have been reddened by dust within the line of sight. 

As a method of assessment for the performance of the fit, the \texttt{SEDFit} code returns the chi-squared metric. We set a maximum $\chi^2$ of 500 as the cutoff to accept the result for a given EB as in the catalog. This threshold was conservatively chosen after visual inspection of a large number of SED fits, where fits with $\chi^2 < 500$ were obviously trustworthy fits. We did not include an EB in the final catalog if the radius of either of the component stars was calculated to be above 0.7 solar radii. This restricted our binaries solely to main sequence M dwarfs. 

\section{Results}\label{sec:results}


In this section, we provide tables and show distributions of properties for binaries included in our catalog. We begin with orbital property distributions in Section \ref{subsec:orb_props}. This is followed by physical property distributions in Section \ref{subsec:phys_props}. We then take a further step in calculating some orbital and physical properties in Section \ref{subsec:calc_props}

A binary is included in the catalog if it satisfied all of the following criteria:
\begin{itemize}
    \item The binary is in the initial FFI EB candidate sample (Section \ref{subsec:sample});
    \item The binary has a T$_{eff}$ $<$ 4200K in the TIC v8.3 (Section \ref{subsec:sample});
    \item The binary satisfies the color cut criteria as specified in Section \ref{subsec:sample};
    \item We have a TGLC LC for the source with no negative values (Section \ref{subsec:data});
    \item The binary successfully passes ephemeris characterization (Section \ref{subsec:ephem_val});
    \item The binary ephemeris is verified as "on target" by LEO (Section \ref{subsec:vet});
    \item The SED fit had to complete successfully and the chi-squared metric of the fit was less than 500 (Section \ref{subsec:sed_fit}).
    \item The SED calculation yields primary and secondary radii which are below $0.7 R_{\odot}$.
\end{itemize}

After all steps, we are left with 1292 \MMs\ in our final catalog. In the following sub-sections, we present our results as tables and distributions. 

\subsection{Eclipse and orbital properties}\label{subsec:orb_props}

\begin{deluxetable*}{cccccccccc}

\rotate
\tablecaption{Eclipse and orbital properties for our \MMs, including orbital periods, mid-eclipse times (epochs), eclipse depths, secondary phase locations, eclipse widths, eccentricities, and effective temperature ratios calculated from eclipse depths. \label{tab:orb_props}}

\tablehead{\colhead{TIC} & \colhead{Period} & \colhead{Epoch} & \colhead{Prim. Depth} & \colhead{Sec. Depth} & \colhead{Sec. Phase} & \colhead{Prim. Width} & \colhead{Sec. Width} & \colhead{Eccentricity} & \colhead{T$_{eff}$ Ratio}\\ 
\colhead{(Identifier)} & \colhead{(days)} & \colhead{(BTJD)} & \colhead{(Rel. Units)} & \colhead{(Rel. Units)} & \colhead{(Phase Units)} & \colhead{(days)} & \colhead{(days)} & \colhead{} & \colhead{}} 

\startdata
316050212 & 0.954149 & 1791.2741 & 0.1514 & 0.1417 & 0.4983 & 0.0484 & 0.0471 & 0.0133 & 0.9837 \\
381805454 & 2.160679 & 1792.9189 & 0.1214 & 0.101 & 0.5017 & 0.09 & 0.0857 & 0.0248 & 0.955 \\
61224047 & 1.249886 & 1601.185 & 0.408 & 0.4072 & 0.5 & 0.0883 & 0.0757 & 0.0768 & 0.9995 \\
186510259 & 2.122556 & 1765.7715 & 0.1031 & 0.096 & 0.5017 & 0.2811 & 0.0806 & 0.0043 & 0.9824 \\
176790779 & 1.792617 & 1764.8137 & 0.2331 & 0.2302 & 0.4983 & 0.1844 & 0.1624 & 0.0634 & 0.9968 \\
464343363 & 5.228863 & 1655.6044 & 0.0829 & 0.0809 & 0.5012 & 0.08 & 0.0805 & 0.0037 & 0.9939 \\
25681445 & 0.482229 & 1765.1717 & 0.4627 & 0.355 & 0.5017 & 0.0638 & 0.0629 & 0.0071 & 0.9359 \\
196902457 & 4.28652 & 1659.1635 & 0.3639 & 0.2936 & 0.4983 & 0.1864 & 0.209 & 0.0573 & 0.9477 \\
55108492 & 6.613529 & 1547.0412 & 0.2545 & 0.1739 & 0.4983 & 0.134 & 0.1335 & 0.0033 & 0.9092 \\
267806373 & 2.135139 & 1766.1757 & 0.0666 & 0.062 & 0.5017 & 0.0524 & 0.0681 & 0.1296 & 0.9823 \\
105767200 & 0.322372 & 1546.1219 & 0.5622 & 0.4809 & 0.4883 & 0.0695 & 0.0613 & 0.0652 & 0.9617 \\
273263945 & 10.39735 & 1816.0204 & 0.2879 & 0.1625 & 0.545 & 0.1242 & 0.1145 & 0.0815 & 0.8669 \\
459083368 & 0.387406 & 1816.6101 & 0.4974 & 0.4927 & 0.4983 & 0.044 & 0.0432 & 0.0093 & 0.9976 \\
238879024 & 9.599885 & 1818.5635 & 0.5 & 0.5 & 0.5015 & 0.106 & 0.1057 & 0.0027 & 1 \\
27302173 & 2.302155 & 1657.9809 & 0.4149 & 0.384 & 0.4983 & 0.0739 & 0.0716 & 0.0162 & 0.9808 \\
66849985 & 1.281165 & 1816.7356 & 0.1323 & 0.0882 & 0.5017 & 0.0496 & 0.0313 & 0.0905 & 0.9034 \\
339253436 & 2.230005 & 1817.6587 & 0.1242 & 0.1072 & 0.5017 & 0.0745 & 0.0593 & 0.1131 & 0.9638 \\
252304087 & 0.554889 & 1816.3547 & 0.2783 & 0.2169 & 0.5017 & 0.058 & 0.0502 & 0.0719 & 0.9396 \\
244208214 & 1.31432 & 1439.1847 & 0.0639 & 0.0604 & 0.5017 & 0.0488 & 0.0436 & 0.0562 & 0.986 \\
398905942 & 4.113217 & 1439.3584 & 0.259 & 0.1786 & 0.5 & 0.065 & 0.067 & 0.0152 & 0.9113 \\
\enddata

\tablecomments{Table 1 is published in its entirety in the machine-readable format.
      A portion is shown here for guidance regarding its form and content.}
\end{deluxetable*}

\begin{figure}
    \centering
    \includegraphics[width=.95\linewidth]{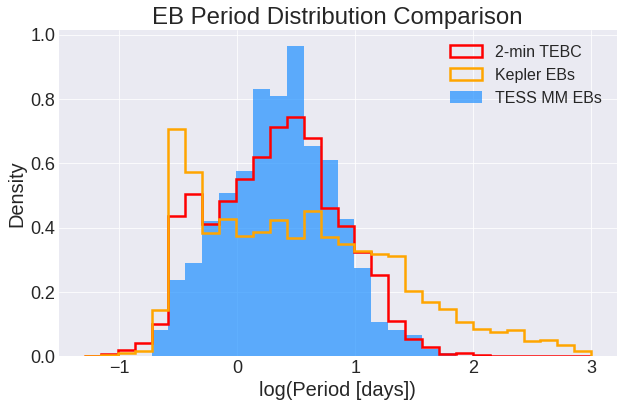}
    \caption{The distribution of the orbital periods of our \MM\ binaries (blue) compared with the \tess\ 2-minute sample (outlined yellow) and the Kepler sample (outlined orange) normalized to the same vertical scale. Orbital period is reported in days and is shown in log-space.}
    \label{fig:period_comp}
\end{figure}

We give our eclipse and orbital properties in Table \ref{tab:orb_props}. In Fig. \ref{fig:period_comp}, we show the distribution of orbital periods for binaries in our final sample in blue. We also compare to the \tess\ 2-minute sample of eclipsing binaries from \citet{Prsa_2022} (outlined in red, TEBC) and the final iteration of the Kepler EB sample from \citet{Kirk_2016} (outlined in yellow, KEBC). Because these catalogs have varying numbers of EBs, we normalize so they are on the same vertical scale in Fig. \ref{fig:period_comp}. While our catalog contains only low-mass \MM\ EBs, both other catalogs contain component stars of varying spectral types. The KEBC sample has the longest tail at the long-period end. This is reasonable given its near-continuous coverage of a small part of the sky for about 4 years. On the other hand, \tess\ is comparatively less sensitive to EB periods beyond about 10 days, which is demonstrated by both our sample and the TEBC sample. While we can leverage the \tess\ observing strategy for sources which have multiple consecutive sectors and sources which were observed in both the PM and EM1, the typical \tess\ source is observed only once every two years. None of these catalogs are representative of the true distribution of binary periods, as the true distribution of binary orbital periods is peaked at much longer periods \citep{ARAA_2013_st_mult}.

\begin{figure*}
    \centering
    \gridline{
    \fig{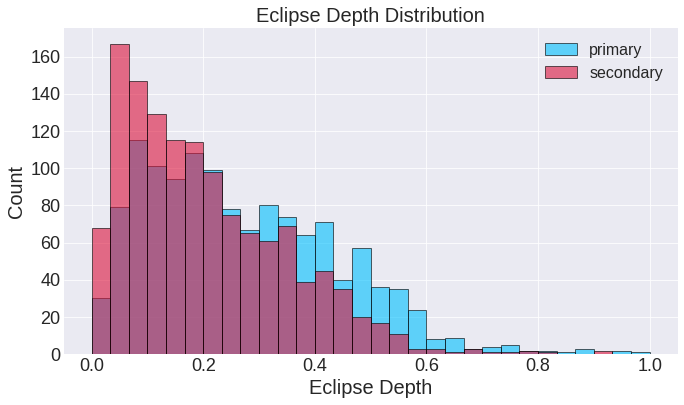}{.48\textwidth}{(a)}
    \fig{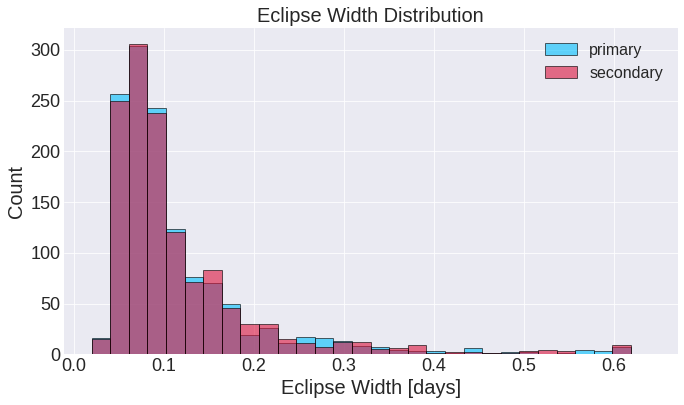}{.48\textwidth}{(b)}
    }

    \gridline{
    \fig{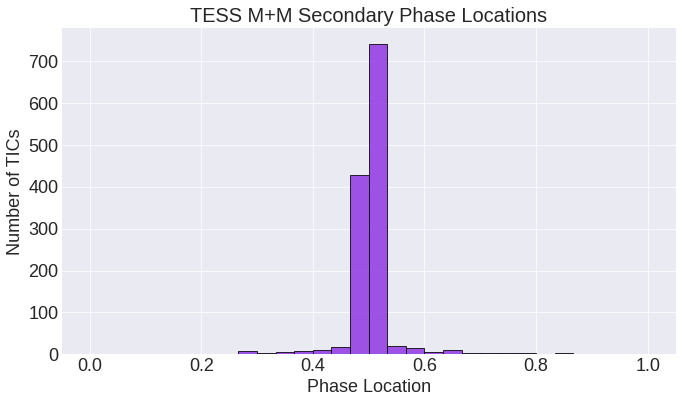}{.48\textwidth}{(c)}
    }
    \caption{Eclipse depth (fractional units), width/duration (days), secondary phase location (phase units) distributions in panels a, b, and c respectively. For the eclipse depths and durations, the primary eclipse (which is defined as the deeper eclipse) is plotted in blue, and the secondary eclipse is plotted in red. The location of the secondary eclipse in phase space relative to the primary eclipse (which is defined to be at phase 0) is plotted in pink.}
    \label{fig:orbital_props}
\end{figure*}

In Fig. \ref{fig:orbital_props}, we show the distributions of eclipse depths, durations, and secondary eclipse phase locations in the top left, top right, and bottom, respectively. Eclipse depths are measured in relative units since light curves are normalized to 1. Eclipse depths go as deep as nearly 100\% in some cases, which is not physical. This is likely due to over-subtraction of the background in these cases. This should not affect the calculated radii of the binary components because only the ratio of eclipse depths is used as an input to the SED fit. Both primary and secondary eclipse depths are primarily concentrated from 0 to about 0.6. Our secondary eclipse depth distribution has a steeper negative slope from 0.05 to 0.6

Eclipse widths are measured in time units of days. Both our primary and secondary eclipse widths are peaked at a duration just below 0.1 days (approximately 2 hours). Our primary and secondary eclipse widths have very similar distributions, with long tails from the peak towards the maximum duration of 0.6 days (14.4 hours).

The location of the secondary eclipse is reported in phase units relative to the primary eclipse, which is defined as being at phase 0. Therefore, a secondary eclipse which is at phase 0.5 is halfway between successive primary eclipses. Our distribution of secondary eclipse phase locations is strongly peaked at 0.5.


\subsection{Physical property distributions}\label{subsec:phys_props}

\begin{splitdeluxetable*}{ccccccBccccccc}

\tabletypesize{\footnotesize}

\tablecaption{Physical properties for our \tess\ \MMs, including dust extinctions, physical radii, effective surface temperatures, surface gravities, and masses, along with system metallicities, separations, and chi-squared metric for the SED fit. \label{tab:SED_props}}

\tablehead{\colhead{TIC} & \colhead{Extinction} & \colhead{Rad1} & \colhead{Rad2} & \colhead{T$_{eff}$1} & \colhead{T$_{eff}$2} & \colhead{Logg1} & \colhead{Logg2} & \colhead{Mass1} & \colhead{Mass2} & \colhead{[Fe/H]} & \colhead{Sep.} & \colhead{ChiSq} \\ 
\colhead{(Identifier)} & \colhead{(Av)} & \colhead{(Sol. Rad.)} & \colhead{(Sol. Rad.)} & \colhead{(K)} & \colhead{(K)} & \colhead{(dex)} & \colhead{(dex)} & \colhead{(Sol. Mass)} & \colhead{(Sol. Mass)} & \colhead{(dex)} & \colhead{(AU)} & \colhead{} } 

\startdata
186510259 & 0.17593 & 0.5113 & 0.4953 & 3802.6 & 3742.0 & 4.1336 & 3.5575 & 0.4565 & 0.4553 & 0.00000 & 0.03134 & 74.50 \\
25681445 & 0.00268 & 0.6602 & 0.6317 & 3946.4 & 3721.6 & 4.5845 & 4.4022 & 0.5734 & 0.5665 & 0.00000 & 0.01257 & 8.35 \\
55108492 & 0.05546 & 0.5526 & 0.4875 & 3639.7 & 3419.1 & 3.8607 & 3.6038 & 0.4983 & 0.4690 & 0.00000 & 0.06819 & 40.66 \\
27302173 & 0.00198 & 0.5993 & 0.5514 & 3781.7 & 3700.0 & 4.0011 & 3.6115 & 0.5434 & 0.5097 & 0.35407 & 0.03471 & 94.68 \\
252304087 & 1.06972 & 0.4883 & 0.4500 & 3706.2 & 3439.9 & 6.2673 & 3.2236 & 0.4810 & 0.4720 & 0.00000 & 0.01300 & 45.64 \\
398801705 & 0.13887 & 0.5368 & 0.5368 & 3525.0 & 3525.0 & 4.0000 & 4.0000 & 0.4992 & 0.4968 & 0.20853 & 0.01236 & 65.08 \\
354535677 & 0.13374 & 0.5573 & 0.3323 & 3733.5 & 2715.2 & 5.5224 & 3.9480 & 0.5515 & 0.3216 & 0.00000 & 0.02983 & 48.13 \\
87728707 & 0.32606 & 0.5118 & 0.4489 & 3853.3 & 3526.7 & 4.3878 & 3.8195 & 0.4716 & 0.4116 & 0.00000 & 0.10184 & 33.36 \\
426219178 & 0.78684 & 0.5609 & 0.5067 & 3700.0 & 3693.4 & 5.8512 & 4.0019 & 0.5263 & 0.4748 & 0.17046 & 0.03259 & 36.36 \\
283549399 & 0.00035 & 0.4042 & 0.1742 & 3439.2 & 2700.8 & 5.1848 & 2.4098 & 0.3714 & 0.1212 & -0.04851 & 0.01588 & 36.85 \\
425157052 & 0.00000 & 0.5741 & 0.4754 & 3949.6 & 2982.8 & 4.4993 & 3.6357 & 0.5454 & 0.4408 & 0.00000 & 0.01753 & 57.84 \\
58019376 & 0.28614 & 0.6131 & 0.5468 & 3910.1 & 3360.7 & 5.4842 & 3.0000 & 0.5532 & 0.5138 & -0.10873 & 0.06058 & 296.87 \\
353067141 & 0.24798 & 0.5932 & 0.5438 & 3937.3 & 3156.4 & 5.5709 & 3.9171 & 0.5380 & 0.5293 & 0.00000 & 0.01875 & 5.19 \\
148442084 & 0.45539 & 0.6111 & 0.4278 & 3529.2 & 1874.8 & 4.6331 & 4.1778 & 0.5400 & 0.3961 & 0.00000 & 0.03135 & 9.71 \\
28754926 & 0.33688 & 0.4873 & 0.1573 & 3600.0 & 2542.7 & 5.4137 & 2.5023 & 0.4536 & 0.0877 & -0.27430 & 0.08218 & 35.40 \\
155295440 & 0.00000 & 0.5290 & 0.5290 & 3611.9 & 3611.9 & 3.9337 & 3.9337 & 0.5174 & 0.5113 & 0.00000 & 0.05013 & 46.95 \\
61505537 & 0.15113 & 0.5948 & 0.5318 & 3678.9 & 3255.6 & 7.2215 & 3.0000 & 0.5538 & 0.4852 & 0.33376 & 0.03962 & 37.70 \\
200245885 & 0.01735 & 0.5511 & 0.5443 & 3838.9 & 3618.7 & 4.4721 & 4.3533 & 0.5165 & 0.5103 & 0.00000 & 0.09691 & 22.94 \\
334155990 & 0.25046 & 0.5934 & 0.2819 & 3960.4 & 3454.4 & 4.5000 & 4.0000 & 0.5468 & 0.2886 & -0.23038 & 0.01538 & 51.45 \\
174830869 & 0.16617 & 0.5617 & 0.5617 & 3983.9 & 3983.9 & 3.0000 & 3.0000 & 0.5154 & 0.5047 & 0.13253 & 0.01102 & 25.44 \\
\enddata

\tablecomments{This table is published in its entirety in the machine-readable format. A portion is shown here for guidance regarding its form and content.}

\end{splitdeluxetable*}

We ran the SED fit code for 4675 candidate \MM\ EBs, which were the output of the ephemeris verification and vetting steps. Of those, 3525 candidates (75.27\%) completed successfully the SED fit code, meaning that more than 1000 sources did not successfully finish the SED fit. Most often, a failure to complete the SED code was due to two primary reasons: 1. There was either no Gaia parallax for the source or the Gaia parallax error was greater than 20\%; 2. The sampler for the SED fitter stepped outside the bounds of the stellar atmospheric models. Of those 3,525 sources that successfully completed the SED code, only the 1,292 in the final catalog satisfied all conditions specified at the beginning of Section \ref{sec:results}.

We use the \texttt{SEDFit} results to report physical properties of the component stars of our \tess\ \MM\ EBs, including the effective temperatures (in Kelvin), physical radii (in solar radius units), surface gravities ($log(g)$), and masses (in solar mass units), and system metallicity ([Fe/H]). We also report magnitudes of dust reddening in the V band (Av) and chi-squared metric for the goodness of fit in Table \ref{tab:SED_props}. 

\begin{figure*}
    \centering
    \gridline{
    \fig{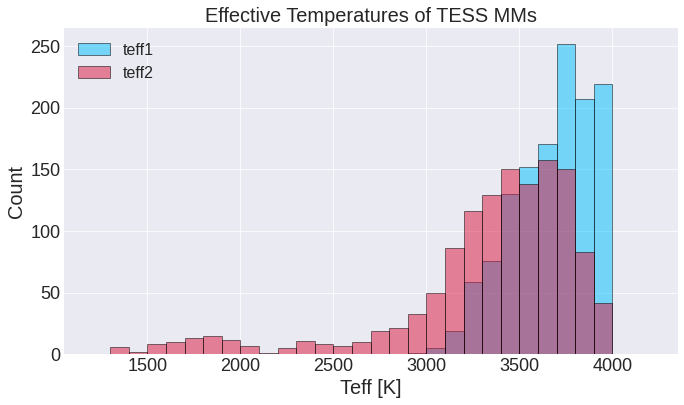}{.48\textwidth}{(a)}
    \fig{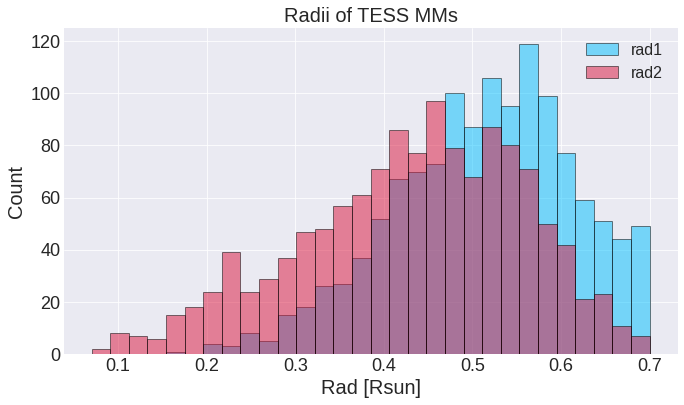}{.48\textwidth}{(b)}
    }

    \gridline{
    \fig{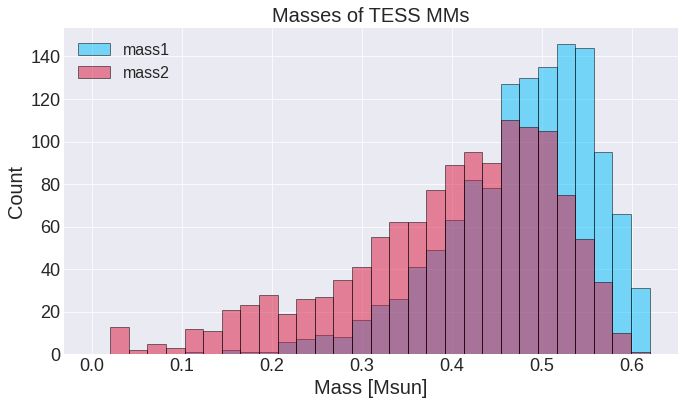}{.48\textwidth}{(c)}
    }

    \caption{Calculated physical properties of the component stars, including effective temperature, physical radii, and masses in panels a, b, and c, respectively. For all panels, the primary star properties are plotted in blue and the secondary star properties are plotted in red.}
    \label{fig:phys_props}
\end{figure*}

In Fig. \ref{fig:phys_props}, we show the physical properties of the component stars from our SED fits, including effective temperatures, physical radii, and masses in the top left, top right, and bottom, respectively. 
These panels all show that the components of the stars in our catalogs are chiefly early M dwarfs (i.e. M0 - M3). This is likely the result of bias towards early M dwarfs. Our catalog does not contain many primaries below 0.35 $M_\odot$, so it is not possible to split the sample across the radiative-convective boundary in a statistically-significant way.
For all three of the plotted properties, the properties of the primary stars are consistent with M-dwarf spectral types. However, for all three of the plotted properties, some of the secondaries have properties which would be consistent with sub-stellar companions at the lowest bound of the grid used to compute stellar atmospheric models. 

\subsection{Calculated Properties}\label{subsec:calc_props}
Using the orbital and physical properties from the previous sub-sections, we can calculate properties which give us better insights into the binary population in our catalog. 

\subsubsection{From eclipse properties}
Using the orbital properties as discussed in Section \ref{subsec:orb_props}, we can calculate properties such as orbital eccentricities and effective temperature ratios, which are included as columns in Table \ref{tab:orb_props}.

Effective temperature ratio can be calculated from eclipse depths. This estimation is based on a simple assumption that both stars in the binary can be approximated as circular discs of a given blackbody temperature which are occulting one another. Limb darkening and line-of-sight geometry are not taken into account in this approximation. Effective temperature ratio calculated in this way was used as an input to the SED fit, and we can then compare effective temperature ratios from eclipse depths and from SED fits. 

\begin{figure}
    \centering
    \includegraphics[width=0.95\linewidth]{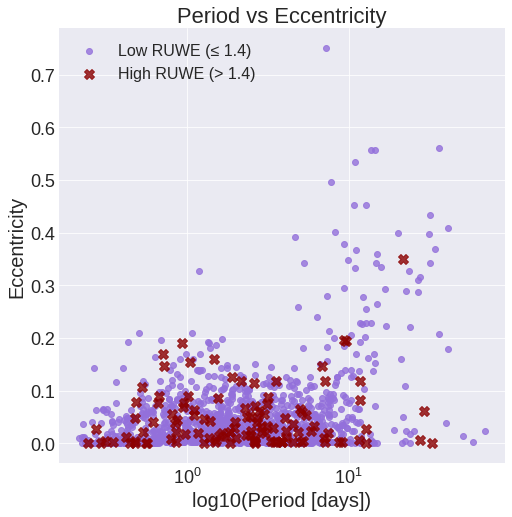}
    \caption{Orbital periods vs eccentricities for our sample. Points are distinguished by both shape and color into high and low RUWE sub-samples by making a cut at RUWE = 1.4.}
    \label{fig:per_ecc}
\end{figure}

We calculate orbital eccentricities from secondary phase locations and eclipse durations by using equations 1-3 in \citet{Martin_2021} and by using the trigonometric identity $e = \sqrt{(e\cos\omega)^2 + (e\sin\omega)^2}$. The distribution of orbital period vs eccentricity is shown in Fig. \ref{fig:per_ecc}. A majority of binaries in our sample have very low eccentricities below 0.1, which is to be expected. This is expected because tidal circularization is strong at short orbital periods and decreases significantly beyond orbital periods of approximately 10 days.

\subsubsection{From SED properties}\label{subsubsec:phys_props}


    

We can use the properties from our SED fit to calculate properties such as mass ratios, effective temperature ratios, and binary separations, though only binary separation is included as a column in Table \ref{tab:SED_props}. The mass ratio and effective temperature ratio are calculated as above (e.g. mass ratio $q = \frac{m_2}{m1}$ where $m_{1,2}$ are the masses of the primary and secondary). 

\begin{figure}
    \centering
    \includegraphics[width=0.95\linewidth]{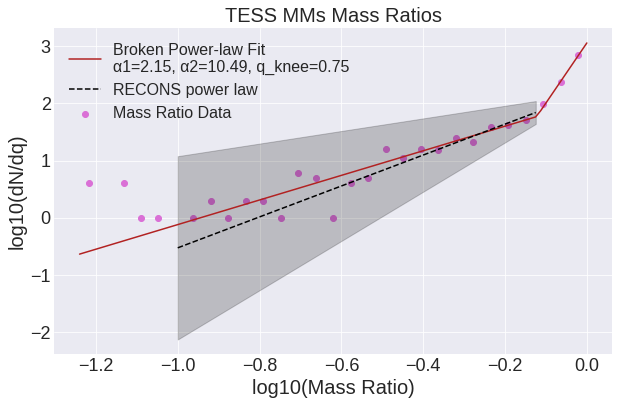}
    \caption{Mass ratios with a fitted broken power law, as in Equation \ref{eqn:mrat_powerlaw}. We also compare to the \citet{ARAA_2013_st_mult} fit of RECONS data \citep{Henry_2006} with errorbars (shown as the shaded region), which is in agreement with our power law index.}
    \label{fig:mrat_powerlaw}
\end{figure}

While power laws ($dN / dq \propto q^a$) are imperfect for describing the complex behaviors that shape binary mass ratio distributions, they are an interesting method for comparing 
this property for different primary mass ranges (e.g. \citet{DeFurio_2022}). While the assumption of random pairing might indicate that binary mass ratios would fit to a Salpeter initial mass function (IMF) at the high-mass end (exponent $a$ of about $-2.3$) and increase to an exponent of $a\sim 0$ (i.e. flat distribution) or below \citep{Tout_1991}, this has not yet been confirmed, and real mass ratio distributions sit well above this \citep{ARAA_2013_st_mult}. Using data from \citet{Henry_2006}, \citet{ARAA_2013_st_mult} find that $a \approx 2.7\pm 1.6$ for short-period M-dwarf binaries from the RECONS sample, but M-dwarf binaries across a broader mass and separation range fit to a power law with $a \sim 0.4$ \citep{Delfosse_2004}. We choose to fit our mass ratio distribution as a broken power law, so as not to bias our fit. Our broken power law takes the functional form as the following:

\begin{equation}\label{eqn:mrat_powerlaw}
    \frac{dN}{dq} = \begin{cases} b_1 \cdot q^{a_1}, & \text{if } q < q_{\text{knee}} \\ b_2 \cdot q^{a_2}, & \text{if } q \geq q_{\text{knee}} \end{cases}
\end{equation}

Where:
\begin{itemize}
    \item $a_1$: Exponent of the power law for $q < q_{\text{knee}} $;
    \item $a_2$: Exponent of the power law for $q \geq q_{\text{knee}} $;
    \item $b_1$: Normalization constant for the first segment;
    \item $q_{\text{knee}}$: The location of the break-point (i.e. "knee") where the power law transitions;
    \item $b_2 = b_1 \cdot q_{\text{knee}}^{a_1 - a_2}$ ensures continuity at $q_{\text{knee}}$, but is not fitted.
\end{itemize}

In Fig. \ref{fig:mrat_powerlaw}, we show our binary mass ratios scaled to equally-spaced logarithmic bins. The broken power law fit is over-plotted as well. We find that our break-point is at $q_{\text{knee}} = 0.75\pm0.02$, with exponents above and below the break-point as $a_1 = 2.15\pm0.38$ and $a_2 = 10.49\pm0.37$, respectively. Our large exponent for high-mass ratios (above the break-point at $q = 0.75$) further demonstrates the propensity for so-called twin binaries in our sample. However, the exponent $a_1$ below the break-point is consistent with the fit to the RECONS sample within uncertainties, which is also plotted in Fig. \ref{fig:mrat_powerlaw}. This consistency may indicate that our photometric survey and their spectroscopic survey are probing a similar population of nearby low-mass M-dwarf binaries.

\section{Discussion}\label{sec:disc}

We organize this section into five main parts, choosing to discuss some of the most important takeaways and implications from this work. 

\subsection{Contact binaries and the shortest-period \MMs\ }

In Fig. \ref{fig:period_comp}, our period distribution has a shape which is similar to the TEBC sample, with an exception in the short period regime. Both the TEBC and the KEBC samples exhibit an excess of very short-period binaries with orbital periods below 1 day. \citet{Prsa_2022} attribute this feature to a sub-sample of contact binaries\footnote{There is some discussion in the literature regarding the subtleties between the words ''contact" and ''overcontact". While the term ''contact binary" might refer to one in which both components overflow their Roche lobes, ''overcontact" might imply frequent mass exchange or even an overlap in stellar surfaces or other structures. We do not make such a distinction and only use the term ''contact" binaries.}. Both the TEBC and the KEBC contain a high fraction of solar-type binaries. While we should be sensitive to short-period \tess\ EBs \citep{Kostov_2025}, this difference may indicate that \MMs\ in the shortest-period orbits are more rare compared to solar-type binaries.

What is the shortest-period orbit that can be sustained by a main sequence binary? Previous literature has suggested that the short-period limit is approximately 0.22 days (e.g. \citet{Rucinski_1992}). This limit has been physically motivated by the timescales on which angular momentum loss (AML) from a detached configuration to contact occurs. The core argument is that such AML does not often occur within the Hubble time \citep{Stepien_2011}. In other words, if we had the time to wait, we may later see a larger proportion of binaries below this period threshold. As an additional mechanism explaining the dearth of such short-period binaries, \citet{Jiang_2012} argues that mass transfer between the binary components is unstable at such short orbital periods, which will lead to rapid merger events. 

Binaries at orbital periods below this limit - including low-mass M dwarf binaries - have been observed (e.g. \citet{Nefs_2012,Davenport_2013,Qian_2015,Soszynski_2015,Xu_2022}). Finding M dwarf binaries below this limit has multiple explanations: 1. AML timescales are overestimated \citep{Nefs_2012}; 2. M dwarf binaries form differently from solar-type stars and are subject to different AML timescales \citep{Soszynski_2015}; 3. Orbits are shrunk beyond the 0.22 d limit by Kozai-Lidov oscillations in the presence of a tertiary companion \citep{Xu_2022}. 

The shortest-period binary in our sample is 0.215 d (TIC 422884542), which is very close to the proposed short-period limit at 0.22 d. This is the only source in our catalog which has an orbital period below 0.22 d (the next-closest is TIC 57612025 at 0.225 d). Both the TEBC and KEBC contain entries with orbital periods below 0.1 days, easily surpassing this limit. Given that we do not find a substantial number of the shortest-period binaries, this might indicate that M dwarf binaries have not had time to lose more angular momentum to further shrink their orbits (scenario 2 above), and that binaries with shorter orbital periods must have been subjected to some external process that further shrunk their orbits (scenario 3 above). Alternatively, this could also mean that such binaries tend to merge at periods below 0.22 days. 

The dearth of \MM\ EBs below 0.22 d in this catalog in combination with the missing contact binaries is informative for very low-mass binary evolution. Our period distribution in comparison to the TEBC and KEBC samples might suggest that short-period \MMs\ are either more quickly destabilized and merge at periods below 1 d or are orbitally supported at periods beyond 1 d. While we must be cautious to draw explicit population-level conclusions in the absence of a complete catalog, these results provide further context for how short-period low-mass binaries evolve.

\subsection{Eccentricities and calculating a circularization period}

Our period-eccentricity plot in Fig. \ref{fig:per_ecc} tells a rich dynamical history of these binaries. We see a similar envelope shape which has been observed in other similar studies \citep{Zanazzi_2022,Vrijmoet_2022}, including non-zero eccentricities even at very short orbital periods. This may be lasting evidence of dynamical interactions from the formation of short-period binaries, such as with circumbinary disks of material prior to the dissolution of such disks \citep{Artymowicz_1991,Siwek_2023,Valli_2024}. We also see a small number of binaries in our sample which still have circular orbits even at longer orbital periods relative to others in the sample, as seen in the lower right corner of the period-eccentricity plot in Fig. \ref{fig:per_ecc}. Therefore, any mechanism invoked to describe the formation and migration of low-mass binaries to their current orbits must be able to describe a distribution of eccentricities, including both very low eccentricities moderately excited orbits.

Such a large sample of short-period eclipsing binary systems yields insights into the circularization period of these binaries. The circularization period refers to the orbital period below which nearly all systems are found to have circular orbits and is a probe of the efficiency of tidal dissipative forces \citep{Meibom_2005,Zanazzi_2022}. Circularization period manifests at population level and is expected to change as a function of system age and will also scale with primary mass and the radii of both of the components. It has been suggested that the circularization period for a group of EBs can be used to find the age of the population \citep{Mathieu_1988}. Previous literature has suggested that at periods shorter than 7 days, low-mass binaries are strongly tidally circularized \citep{Vrijmoet_2022}, which is shorter than the circularization period for solar-type binaries at about 12 days \citep{Raghavan_2010}. More massive primaries tend to have shorter circularization periods than solar-type binaries (e.g. \citet{Bashi_2023,IJspeert_2024}).

We calculated the circularization period of this sample in a novel statistical way. To find the circularization period for our sample, we repeatedly split the sample into ''short period" and ''long period" groups many times on a period scale from 1 to 10 days and then performed a Kolmogorov-Smirnov (KS) test on the eccentricities of the low and high period subsamples to see whether the samples are drawn from the same distribution. Based on this analysis, the p-value is strongly peaked at a period cut of approximately 2.0 days, meaning there is an inflection point in our eccentricity distribution at this period. We argue that this may be representative of the population-level circularization period for these low-mass EBs. 

A circularization period of 2.0 days for this population is shorter than both the circularization periods of other Kepler and \tess\ solar-type EBs at $\sim3$ days \citep{Zanazzi_2022} and of M-dwarf binaries at $\sim7$ days suggested by \citet{Vrijmoet_2022}. A natural explanation for the circularization period manifesting at this period is equilibrium tide dissipation by convective turbulence (see \citet{Zahn_1989} and similar). Further, dynamical tides supported by resonance locking might also circularize solar-type binaries out to $\sim3-4$ days \citep{Zanazzi_2021}, and it may be the case that dynamical tides are more effective for lower-mass binaries, thus shrinking the circularization period. Many of these processes occur in a timely fashion during pre-MS and early MS phases of stellar evolution. Interior structure and stellar mass also play a crucial role. Because this problem is multi-dimensional along stellar mass and age axes, there is still much more work to be done to reveal the circularization processes of low-mass binaries. In particular, a more fine-grained analysis of circularization period as a function of system age across the stellar mass scale is warranted.

\subsection{\MM\ binaries as twins}

The steep positive slope in our mass ratio distribution towards high mass ratios in Fig. \ref{fig:mrat_powerlaw} is evidence for a strong prominence of so-called ''twin" binaries (i.e. those with mass ratios above q = 0.95). Multiple studies have consistently shown an excess of twin binaries across a range of separations and mass scales (\citet{Simon_2009,Kounkel_2019,ElBadry_2019} and references therein), including late-type M dwarfs \citep{Janson_2014}. While it has been suggested that such prominence of twins could be the result of observational bias \citep{Cantrell_2014}, there could be physical reasons motivating the formation of more twin binaries than other types of binaries. Twin binaries could be formed via competitive accretion in a common disk \citep{Tokovinin_2000} and by disk fragmentation and migration \citep{Tokovinin_2020}. It may be the case that more equal-mass binaries are expected in the low-stellar mass regime \citep{Nefs_2013}, which is consistent with the findings of our catalog. More confounding, however, is the finding that upon closer inspection of a subset of "twin" binary systems, the components do not have similar activity indicators, suggesting that twin formation and evolution histories are difficult to predict \citep{Couperus_2025}.

It is notable that our method is likely to be especially biased towards twin binaries. There are multiple cases in which binaries of varying mass ratios could be errantly characterized as twin binaries in our case. For example, binaries which are geometrically singly eclipsing would be characterized as twin binaries with twice the true period in our catalog. Further, shallow secondary eclipses below the noise threshold might be missed and the binary might be cataloged as equal-mass. Broadly speaking, signal-to-noise for a periodic signal is related to the depth of the signal, the noise in the light curve, and the number of eclipses which can be co-added to increase signal. 

\begin{equation}\label{eqn:SNR}
    SNR \propto \sqrt{N_{ecl}}\frac{\delta}{\sigma}
\end{equation}

As we've shown in Fig. \ref{fig:Tmag_CDPP}, noise is related to magnitude, which itself is a function of both distance and system luminosity. We are more sensitive to eclipses in brighter systems where point-to-point scatter is lower. The number of eclipses is related to the orbital period, where shorter period orbits yield more eclipses in a light curve which can be co-added, thus increasing signal. Therefore, we are more sensitive to shorter-period EBs with deeper eclipses. For a small number of eclipses, SNR in equation \ref{eqn:SNR} approaches 1 as the depth of the eclipse becomes comparable to the noise level. The median \tess\ magnitude in our catalog is quite dim at T = 14.23 mag, with a median CDPP of $3.32 \times 10^{-3}$. We can equate the eclipse depth to this CDPP value, representing an extreme case where our SNR threshold is 1, and there is only one secondary eclipse to detect. In this case, the corresponding eclipse depth ratio is approximately 0.05. Our detection threshold is likely higher than this, representing a minimum detectable eclipse threshold. Binaries with eclipse depth ratios around or lower than 0.05 will likely be mis-categorized as equal-depth binaries. Conversely, this will lead to an over-count of binaries in the highest eclipse-depth ratio bin above 0.95. 

\subsection{Additional Companions to our EBs}\label{subsec:companions}
There is previous evidence to suggest that many low-mass short-period binaries are a part of higher-order systems \citep{Borkovits_2016}. This could indicate that the shortening of the periods of low-mass eclipsing binaries is often facilitated by the presence of a tertiary companion, mediated by angular momentum exchange through Kozai-Lidov oscillations. 
There are multiple lines of evidence that could reveal the existence of additional companions to our EBs including high-resolution imaging, spectroscopic followup, common proper motion analysis, astrometric analysis, eclipse timing variations (ETVs), additional transit/eclipse signals in our light curves, and more. These various lines of evidence probe different separation and mass scales, and must be used in tandem to paint a more complete picture of the prospects for additional companions. Additionally, the completeness/sensitivity of the various methods should be assessed. Such efforts would be revealing for the ways in which low-mass short-period binary systems form and evolve. 

Various choices we made along the way may have biased our sample towards binaries that contain only main sequence components. Similarly, these same choices may have impacted the kinds of companions we might expect to find orbiting our binaries since the construction of the catalog was not a fully random search for \MMs. Based on our SED fitting, we may not expect to find additional luminous companions in short-period orbits, but future analysis could yield luminous companions in widely separated orbits. Such companions might be revealed through common proper motion analysis.

\subsubsection{Gaia Astrometry}
\begin{figure}
    \centering
    \includegraphics[width=.9\linewidth]{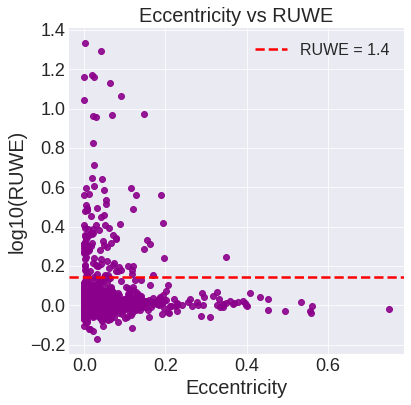}
    \caption{\MM\ EB orbital eccentricity vs Gaia DR3 RUWE (plotted in log space). We place a dashed red line at RUWE = 1.4, indicating our "high RUWE" cutoff.}
    \label{fig:per_ecc_ruwe}
\end{figure}

The Gaia DR3 archive gives us parallax distances and Renormalized Unit Weight Error (RUWE) for the astrometric solutions used to calculate distance. Typically, it is assumed that a RUWE  $\approx 1$ is a good astrometric solution for a single star, whereas a RUWE $>1.4$ is due to a spurious solution and may indicate that the source is part of a separated binary or higher order pair, though this is not the only interpretation of high RUWE for a given system. Gaia measures angular perturbations of the photo-center of a source, so converting to a physical size scale is distance-dependent. This means that nearer sources can yield higher RUWE for the same binary separation, and even sources with separations as small as 0.1 AU might show evidence of higher RUWE \citep{Belokurov_2020}. 

In Fig. \ref{fig:per_ecc_ruwe}, we show log(RUWE) plotted against eccentricity. There appears to be a transition in the RUWE distribution at an eccentricity of 0.2, where the RUWE distribution narrows significantly at high eccentricity. This offers a slightly counter-intuitive picture to the relationship between RUWE and eccentricity. Systems with moderate eccentricities might have been dynamically excited by a massive tertiary companion, and high RUWE might be suggestive of a tertiary companion, so we might expect the two to be correlated. However, the discrimination between high-eccentricity and high-RUWE systems might mean that we are probing two different types of tertiary companions for this population of binaries. High RUWE might be caused by a massive luminous and widely separated companion, so these systems might be good targets to search for stellar companions. Moderate eccentricities might be caused by dynamical interactions of a tight binary with a circumbinary disk, from which circumbinary planets might form, so systems with moderate eccentricity might be good targets to search for sub-stellar companions.

In order to untangle the dual impacts of our binaries and additional companions on RUWE, we can calculate the expected RUWE using equations from \citet{Belokurov_2020}. The expected RUWE scales with distance to the system, separation of the binary, and binary mass and luminosity ratio. A higher RUWE is expected for a system which is nearer, more widely separated, and/or has a lower mass/luminosity ratio. We found that anomalously-high RUWE as a result of the inner binary itself is not expected for a vast majority of our systems, save for three examples (TIC 117378460, TIC 148675222, and TIC 318057648). These three systems are relatively close, widely separated, and have relatively low mass ratios. Any other system which has a high RUWE might have a tertiary companion \citep{Stassun_2021}, but we caution against strictly interpreting RUWE in this way. High RUWE is not necessarily due to the presence of a tertiary companion, and low RUWE does not necessarily imply that a system does not have a tertiary companion. 

For statistical comparison, we split the full catalog into a "low RUWE" sample and a "high RUWE" sample by creating a cutoff at RUWE = 1.4. The high RUWE sample TICs are listed in Table \ref{tab:hi_ruwe}. We checked whether the low and high RUWE sample period and eccentricity distributions were statistically different with KS tests. For both period and eccentricity, we found that the low and high RUWE samples are closely related with very high p-values above 0.9 for both parameters. This again shows that RUWE does not explicitly correlate with the presence of tertiary companions, but may instead serve as a suggestion for the first systems to check.

\subsection{Implications for binary star formation and evolution}

With this catalog, we present a sample of low-mass, short-period binaries comprised of main sequence M dwarfs which tend to have similar masses in circular-to-moderately-eccentric orbits. These binaries are likely described by turbulent fragmentation, followed by a period of migration to very short-period orbits, during which time it is hypothesized that preferential accretion onto the secondary equalized the masses of the components in a majority of cases. These binaries have likely undergone competing processes of tidal circularization and eccentricity excitation through circumbinary disk interactions, imprinting them with a range of eccentricities. There is clearly variation on this theme within this catalog, and there are many more factors to consider, including formation environments, system ages, component interior structures and activity cycles, and others.  

Tertiary companions may play a crucial role in determining the evolutionary histories of these multiple systems. If it is later revealed that there is a large majority of short-period M dwarf binaries which have tertiary companions, this will have profound implications for how such systems form. A large proportion of tertiary companions to short-period low-mass binaries would naturally explain the shortening of their orbits. It will be especially telling if there is a substantial difference in the fraction of close M dwarf binaries and close solar-type binaries with tertiary companions. 
This will also be informative for the presence - or absence - of CBPs in such systems. Much of the work in simulating CBP formation and evolution has simulated short-period binary systems at the separations we see them now with a CB protoplanetary disk of material. However, we must seek to build a holistic picture of short-period binary star and CBP formation in order to fully encapsulate the stories of these planetary systems. 

\section{Future Directions \& Conclusions}\label{sec:conc}

\subsection{Future work}

In our catalog, we've presented a characterization of the orbital and physical properties of these binaries using fairly simple periodicity and SED analyses. We made these choices due to the volume of \tess\ EB candidates and in an effort to maintain homogeneity where possible. There are many avenues for future work with this catalog. We propose two major avenues here. 

First, in order to test and compare our methodology, it would be advisable to follow up a number of these EBs with spectroscopic instruments and surveys. The physical properties of the components of the binaries could be recalculated with these data and compared to the properties reported in this catalog. Such detailed spectral characterization was not reasonable for this catalog but is warranted in order to illuminate the physical properties of these binaries in greater detail.

Second, a more holistic picture of the presence (or absence) of tertiary companions to these binaries must emerge. This can occur through more detailed astrometric analysis, high-resolution imaging, ETV analysis, or other methods. We propose that future work pursue these directions for this catalog.

\subsection{Conclusions}

We have presented our catalog of 1292 \tess\ \MM\ EBs, including the steps we took to select our sample, calculate ephemerides, vet their signals, and calculate physical properties from broadband SED analysis. We have compared the distributions of the properties for these \MMs\ to other catalogs where appropriate. The findings of this work can be summarized as follows:

\begin{itemize}
    \item We are sensitive to very short-period EBs, with a peak of our period distribution around 3.4 days;
    \item We did not find evidence of the same sub-population of contact binaries found in similar samples of EBs comprised mainly of solar-type binaries \citep{Prsa_2022}. We do not find EBs with periods below 0.22 days (with one sole exception just below this threshold). Together, these attributes of our sample may suggest that \MMs\ are either more quickly destabilized at very short orbital periods, or have not had sufficient time to reach shorter orbital periods;
    \item We found that our short-period binaries tend to be tidally circularized, which is expected for such short-period systems. We calculated that the circularization period of our sample of binaries is around 2.0 days, which is comparable to that found by \citet{Zanazzi_2022} for other Kepler and \tess\ EBs, but is shorter than that found by \citet{Vrijmoet_2022} for nearby M dwarf systems; 
    \item We found a strong proclivity for "twin" EBs with high mass ratios within our catalog, which is consistent with other samples. We further found that the slope of the mass ratio distribution below mass ratios of 0.75 is in agreement with the RECONS sample of low-mass binaries \citep{Henry_2006,ARAA_2013_st_mult};
    \item We briefly explored the possibilities for tertiary companions to these binaries, which will have implications for the formation and evolution of low-mass short-period binaries. RUWE does not strictly correlate with the presence of tertiary companions, so more work is needed to this end.
\end{itemize}

\section{Acknowledgments}.

We thank the anonymous referee of this work whose constructive feedback greatly improved the quality and clarity of this work.

D.O. acknowledges support from and is primarily supported by NASA FINESST award number 80NSSC23K1448.

D. D. acknowledges support from the TESS Guest Investigator Program grants 80NSSC23K0769.

V.\,B.\,K. is grateful for financial support from NSF grant AST-2206814 and from NASA grant PSI-1912-SETI .

This research was supported by the UNM Office of the Vice President of Research.

We would like to thank the UNM Center for Advanced Research Computing, supported in part by the National Science Foundation, for providing the research computing resources used in this work.

This paper includes data collected by the \tess\ mission. Funding for the \tess\ mission is provided by NASA's Science Mission Directorate.

We acknowledge the use of public TESS data from pipelines at the TESS Science Office and at the TESS Science Processing Operations Center. Resources supporting this work were provided by the NASA High-End Computing (HEC) Program through the NASA Advanced Supercomputing (NAS) Division at Ames Research Center for the production of the SPOC data products.

This work has made use of data from the European Space Agency (ESA) mission {\it Gaia} (\url{https://www.cosmos.esa.int/gaia}), processed by the {\it Gaia} Data Processing and Analysis Consortium (DPAC, \url{https://www.cosmos.esa.int/web/gaia/dpac/consortium}). Funding for the DPAC has been provided by national institutions, in particular the institutions participating in the {\it Gaia} Multilateral Agreement.


\vspace{5mm}
\facilities{\tess}


\software{astropy \citep{astropy:2013,astropy:2018,astropy:2022},
astroquery \citep{astroquery:2019},
LEO-Vetter \citep{LEO_2024},
SEDFit,
TGLC \citep{Han_2023},
wotan \citep{Hippke_2019}
          }



\appendix



\section{High-eccentricity systems}

\begin{figure*}
    \centering
    \gridline{
    \fig{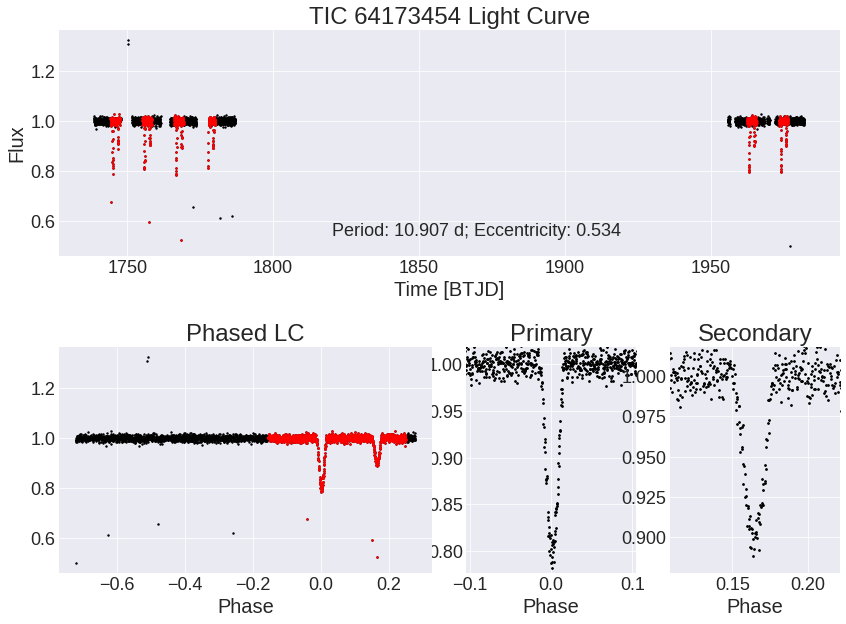}{.48\textwidth}{(a)}
    \fig{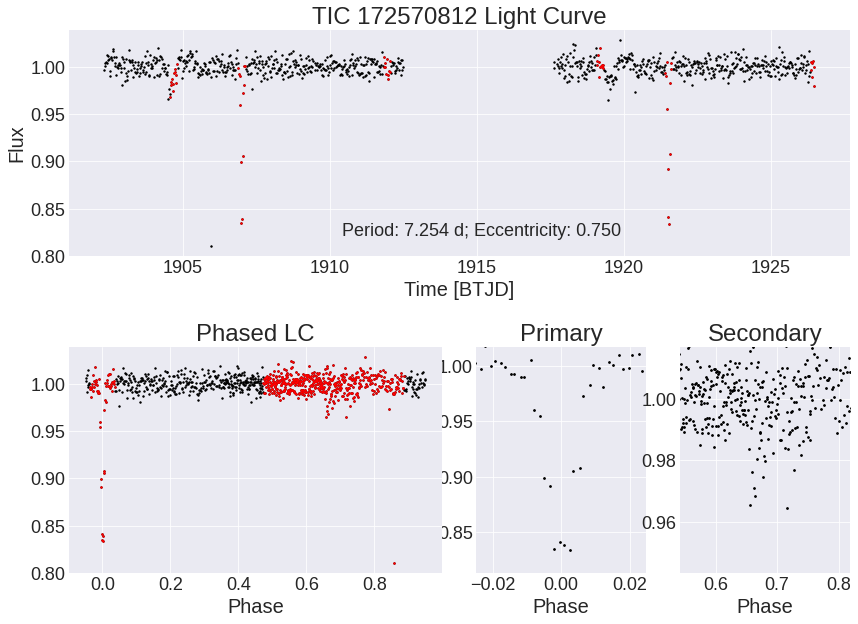}{.48\textwidth}{(b)}
    }

    \gridline{
    \fig{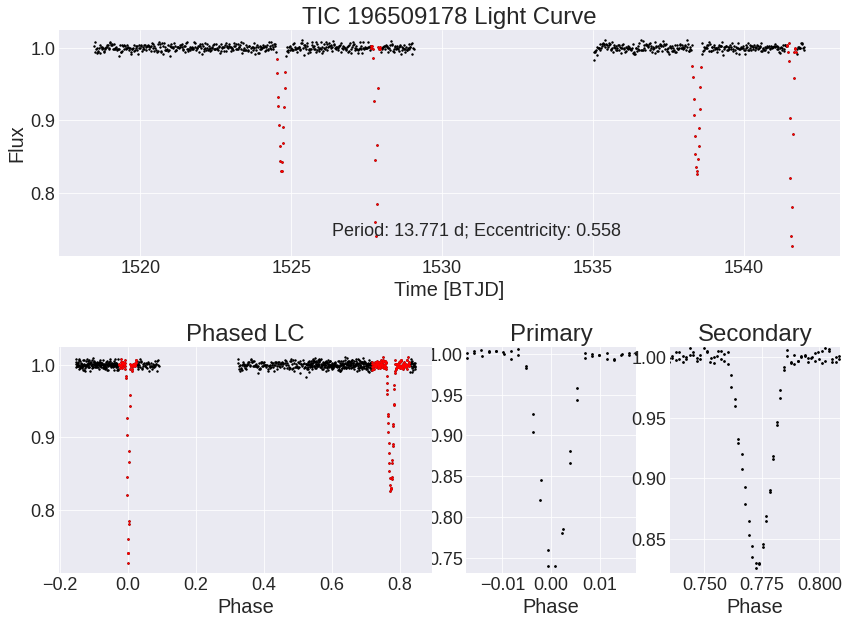}{.48\textwidth}{(c)}
    \fig{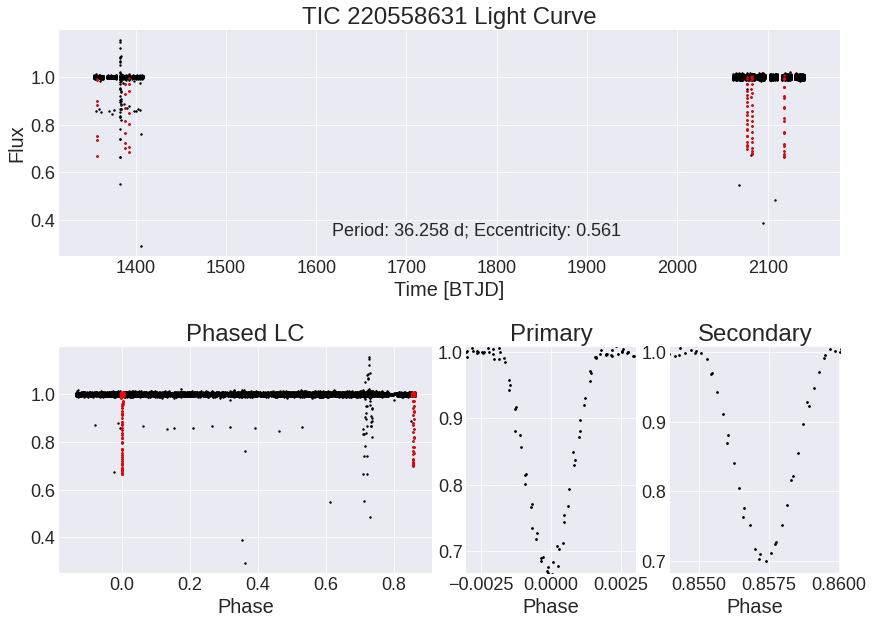}{.48\textwidth}{(d)}
    }
    
    \caption{Light curves, phase-folded light curves, and eclipse profiles for four high-eccentricity systems, including TIC 64173454 (top left), TIC 172570812 (top right), TIC 196509178 (bottom left), and TIC 220558631 (bottom right). In-eclipse points are highlighted in red in the light curve and phase-folded light curve plots.}
    \label{fig:high_ecc_TICs}
\end{figure*}

There are a handful of systems in our sample which have high eccentricities, including eccentricities as high as $e = 0.75$. We have included the light curves, phase-folded light curves, and eclipse profiles of a handful of these high-eccentricity systems in Fig. \ref{fig:high_ecc_TICs} These systems may represent high-value targets for further followup since such high eccentricities at such short orbital periods may be difficult to explain theoretically.

\section{High RUWE systems}

In this Appendix, we present the table of systems with high RUWE.

\startlongtable
\begin{deluxetable}{cccc}\label{tab:hi_ruwe}

\tablecaption{Table of sources in our catalog with Gaia DR3 RUWE $>$ 1.4, representing the ``high RUWE'' subsample.}

\tablehead{\colhead{TIC} & \colhead{Period} & \colhead{Eccentricity} & \colhead{RUWE} \\ 
\colhead{(Identifier)} & \colhead{(days)} & \colhead{} & \colhead{} } 

\startdata
464343363 & 5.228863 & 0.0037 & 1.4969 \\
27302173 & 2.302155 & 0.0162 & 2.0080 \\
28754926 & 11.695962 & 0.0826 & 1.5011 \\
382150036 & 1.107674 & 0.0544 & 3.4423 \\
113355522 & 2.445943 & 0.0522 & 2.5656 \\
166282532 & 0.478491 & 0.0045 & 2.1714 \\
236709277 & 3.156256 & 0.0872 & 2.5995 \\
450344767 & 7.062682 & 0.0104 & 1.9019 \\
271281625 & 2.588197 & 0.1143 & 3.9562 \\
419664745 & 0.843716 & 0.0382 & 1.9151 \\
452698353 & 0.484362 & 0.0780 & 2.1957 \\
446166017 & 3.854322 & 0.0026 & 3.0134 \\
68388891 & 0.534059 & 0.0208 & 2.4487 \\
468603580 & 1.482864 & 0.0026 & 21.5344 \\
117307657 & 4.498827 & 0.0365 & 2.4229 \\
469674899 & 0.853245 & 0.0026 & 2.4839 \\
154578318 & 0.564042 & 0.0000 & 11.0121 \\
335620963 & 7.410942 & 0.0188 & 14.7898 \\
302233773 & 0.310142 & 0.0054 & 1.9059 \\
25419773 & 3.568445 & 0.1180 & 1.4392 \\
130572888 & 9.604731 & 0.1945 & 2.6238 \\
220514552 & 12.730162 & 0.0283 & 1.5958 \\
66573693 & 2.986926 & 0.0566 & 1.9709 \\
465051041 & 2.296551 & 0.0674 & 9.2630 \\
9274044 & 2.498595 & 0.0240 & 5.1554 \\
371383477 & 21.559306 & 0.3500 & 1.7615 \\
47554202 & 3.439354 & 0.0034 & 3.9638 \\
7322727 & 1.905165 & 0.1266 & 3.6440 \\
118162187 & 0.475299 & 0.0000 & 2.0555 \\
157669515 & 0.866844 & 0.0445 & 4.3855 \\
117378460 & 3.141164 & 0.0744 & 1.6304 \\
172496091 & 1.847373 & 0.0000 & 3.6387 \\
162207682 & 0.808202 & 0.0563 & 1.6726 \\
157345367 & 0.478058 & 0.0485 & 2.8634 \\
158824264 & 2.622934 & 0.0000 & 1.8450 \\
7419932 & 0.443425 & 0.0000 & 1.8848 \\
43452110 & 1.267298 & 0.0467 & 2.1199 \\
308266900 & 0.555935 & 0.0000 & 14.5082 \\
198485676 & 0.954624 & 0.0770 & 2.1690 \\
89502636 & 1.753924 & 0.0137 & 2.2899 \\
377731007 & 32.868869 & 0.0000 & 2.0386 \\
199611396 & 0.525905 & 0.1071 & 1.5882 \\
95112238 & 2.723773 & 0.0484 & 2.0316 \\
142409400 & 1.386882 & 0.0421 & 19.5434 \\
393999666 & 1.016197 & 0.0906 & 11.5432 \\
307930725 & 3.888730 & 0.0026 & 1.5046 \\
298163527 & 1.580605 & 0.0136 & 1.4880 \\
140211568 & 0.274053 & 0.0280 & 2.5554 \\
144307820 & 5.083170 & 0.0033 & 2.0099 \\
95057860 & 7.163753 & 0.0026 & 2.0958 \\
241413538 & 0.716733 & 0.1474 & 9.3557 \\
428612328 & 4.857553 & 0.0217 & 2.3193 \\
34077281 & 12.774362 & 0.0000 & 1.9517 \\
366414973 & 4.961940 & 0.0648 & 2.0692 \\
395691612 & 3.543783 & 0.0026 & 2.2891 \\
241841997 & 0.927313 & 0.1901 & 3.6496 \\
242284425 & 11.638478 & 0.0058 & 3.0667 \\
156249398 & 1.551539 & 0.0861 & 1.5662 \\
460055490 & 2.348372 & 0.0504 & 1.4334 \\
171315849 & 1.514510 & 0.0071 & 3.0418 \\
137809567 & 9.400171 & 0.1963 & 1.7292 \\
322740964 & 3.145545 & 0.0337 & 1.5853 \\
253456070 & 4.517903 & 0.0026 & 1.5911 \\
354035282 & 3.030148 & 0.0205 & 1.6304 \\
218983425 & 1.844800 & 0.0224 & 6.6740 \\
351611000 & 0.977425 & 0.0187 & 2.2990 \\
34311799 & 0.614671 & 0.0408 & 2.7933 \\
309679390 & 0.859538 & 0.0212 & 9.1323 \\
220560755 & 1.960634 & 0.0299 & 9.1013 \\
153712855 & 3.396220 & 0.0301 & 3.1220 \\
149862115 & 1.176908 & 0.0192 & 2.0670 \\
376032643 & 2.604397 & 0.0039 & 1.7646 \\
77071778 & 1.643073 & 0.0196 & 4.4462 \\
354672205 & 1.374068 & 0.0101 & 1.8808 \\
77351653 & 1.980702 & 0.0410 & 1.7110 \\
116020737 & 0.351089 & 0.0028 & 1.7135 \\
148675222 & 1.044392 & 0.1550 & 2.1515 \\
257457364 & 2.642529 & 0.0715 & 1.6027 \\
256362692 & 0.668858 & 0.0909 & 1.5033 \\
385791809 & 2.717487 & 0.0412 & 1.6254 \\
102547647 & 5.433333 & 0.0544 & 3.2714 \\
454233941 & 0.706837 & 0.1693 & 1.4201 \\
51443408 & 4.083160 & 0.0241 & 4.0153 \\
363914760 & 27.733197 & 0.0066 & 1.4628 \\
69874547 & 0.794972 & 0.0091 & 2.4393 \\
350960614 & 2.638258 & 0.0076 & 3.1149 \\
454366433 & 0.950716 & 0.0701 & 2.2770 \\
58464534 & 2.800456 & 0.0612 & 1.5415 \\
1541478 & 29.330309 & 0.0619 & 2.4643 \\
137674451 & 3.299853 & 0.0053 & 3.2512 \\
322657565 & 1.282141 & 0.0000 & 2.5642 \\
49053315 & 6.081474 & 0.0324 & 1.4473 \\
71213561 & 2.642634 & 0.0026 & 1.8334 \\
417759809 & 0.293608 & 0.0000 & 2.0403 \\
385143245 & 3.814137 & 0.0047 & 1.4081 \\
39664172 & 8.681675 & 0.0026 & 2.2599 \\
307796463 & 5.869212 & 0.0244 & 14.4294 \\
70490438 & 2.904561 & 0.0229 & 1.7400 \\
73404611 & 5.178008 & 0.0597 & 1.4318 \\
62602104 & 0.421117 & 0.0116 & 3.6597 \\
219707850 & 1.466273 & 0.1609 & 2.0503 \\
454886660 & 0.246355 & 0.0000 & 2.0352 \\
271380828 & 4.040107 & 0.0474 & 3.8460 \\
56473202 & 2.877900 & 0.0307 & 1.5910 \\
423164751 & 6.836416 & 0.1474 & 1.9279 \\
344298995 & 2.467961 & 0.0165 & 2.8351 \\
354518320 & 1.100021 & 0.0638 & 13.4656 \\
88616571 & 4.075140 & 0.0096 & 2.2448 \\
220421583 & 11.754241 & 0.1194 & 3.0928 \\
277992243 & 2.156460 & 0.1193 & 1.5912 \\
250131338 & 0.660856 & 0.0773 & 1.4865 \\
229400580 & 9.134388 & 0.0023 & 1.6080 \\
672717 & 7.181904 & 0.1183 & 1.5094 \\
42099558 & 2.004556 & 0.0102 & 1.9979 \\
\enddata

\tablecomments{Table values have been truncated to 3 decimal places.}


\end{deluxetable}


\bibliography{MM_paper}{}
\bibliographystyle{aasjournal}



\end{document}